\documentclass{aa}
\usepackage{longtable}
\usepackage{multirow}
\usepackage{graphicx,natbib,txfonts,color}
\bibpunct{(}{)}{;}{a}{}{,}

\begin{document}

\title{Post-AGB stars in the SMC as tracers of stellar evolution: the
  extreme s-process enrichment of the 21 $\mu$m star J004441.04-732136.4.
\thanks{based on observations collected with the Very Large Telescope
  at the ESO Paranal Observatory (Chili) of programme number 084.D-0932.} }

\author{K. De Smedt\inst{1}
\and H. Van Winckel\inst{1}
\and A. I. Karakas\inst{2}
\and L. Siess\inst{3}
\and S. Goriely\inst{3}
\and P. R. Wood\inst{2}
}

\offprints{K. De Smedt, kenneth.desmedt@ster.kuleuven.be}

\institute{ Instituut voor Sterrenkunde, K.U.Leuven, Celestijnenlaan 200D,
B-3001 Leuven, Belgium
\and  Research School of Astronomy and Astrophysics, Mount Stromlo Observatory, 
Weston Creek ACT 2611, Australia
\and Institut d'Astronomie et d'Astrophysique, Universit\'{e} Libre de Bruxelles, ULB, CP 226, 1050 Brussels, Belgium
}

\date{Received  / Accepted}

\authorrunning{K. De Smedt et al.}
\titlerunning{Post-AGB stars in the SMC: J004441.04-732136.4}

\abstract
{This paper is part of a larger project in which we want to focus on
the still poorly understood AGB third dredge-up processes 
and associated s-process nucleosynthesis.}
{We confront accurate spectral abundance analyses of
  post-AGB stars in both the LMC and SMC, to state-of-the-art
  AGB model predictions. With this comparison we aim at improving our understanding
  of the 3rd dredge-up phenomena and their dependencies on initial
  mass and metallicity.}
{ Because of the well constrained distance with respect
 to Galactic post-AGB stars, we choose an extra-galactic post-AGB star
 for this contribution, namely the only known 21 $\mu$m object of the
 Small Magellanic Cloud (SMC): J004441.04-732136.4 . We used optical UVES 
 spectra to perform an accurate spectral abundance analysis. 
With photometric data of 
 multiple catalogues we construct a spectral energy distribution and
 perform a variability analysis. The results are then compared to
 predictions of tailored theoretical chemical AGB evolutionary models
 for which we used two evolution codes.}
{Spectral abundance results reveal J004441.04-732136.4 to be one of
  the most s-process enriched objects found
 up to date, while the photospheric C/O ratio of $1.9 \pm 0.7$, shows
 the star is only modestly C-rich. J004441.04-732136.4 also displays a 
 low $[\textrm{Fe/H}]$ = -1.34 $\pm$ 0.32, which is significantly
 lower than the mean metallicity of the SMC. 
From the SED, a luminosity of $7600 \pm 200 \textrm{L}_{\odot}$ is found,
together with E(B-V) = 0.64 $\pm$ 0.02. According to evolutionary post-AGB 
 tracks, the initial mass should be $\approx 1.3 \textrm{M}_{\odot}$. 
 The photometric variability shows a clear period of 97.6 $\pm$ 0.3
 days. The
 detected C/O as well as the high s-process overabundances
 (e.g. [Y/Fe] = 2.15, [La/Fe] = 2.84) are hard to
 reconcile with the predictions. The chemical models also predict a high Pb
 abundance, which is not compatible with the detected spectrum, and a very high
 $^{12}$C/$^{13}$C, which is not yet constrained by observations. The
 predictions are only marginally dependent on the evolution codes
 used.}
{By virtue of their spectral types, favourable bolometric corrections
  as well as their constrained distances, post-AGB stars in external
  galaxies offer unprecedented tests to AGB nucleosynthesis and
  dredge-up predictions. We focus here on one object J004441.04-732136.4,
 which is the only known 21 $\mu$m source
 of the SMC. We show that our theoretical predictions match
   the s-process distribution, but fail in reproducing the detected
   high overabundances and predict a high Pb abundance which is not
   detected. Additionally, there remain
   serious problems in explaining the observed pulsational properties of this source.}


\maketitle


\section{Introduction}\label{sect:intro}

The final evolution of low- and intermediate-mass stars is a fast
transition from the Asymptotic Giant Branch (AGB) over the post-AGB
transit towards the Planetary Nebula Phase (PN), before the stellar
remnant cools down as a White Dwarf (WD). Although this scheme may be
generally acknowledged, there is no understanding from first
principles of different important physical processes that govern these
evolutionary phases. The main shortcomings are related to the lack of
understanding of the mass-loss mechanisms and mass-loss evolution
along the AGB ascent, the subsequent shaping processes of the
circumstellar shells, and the lack of fundamental understanding of the
internal chemical evolution of these stars \citep{habing03, herwig05}.

Here we focus on the the poorly understood AGB 3rd dredge-up
phenomenon, during which products of the internal nucleosynthesis are
brought to the surface of the star.  This is mainly $^{12}$C as
the primary product of the triple alpha reaction, but also the products of
neutron-capture synthesis.
There are two main neutron sources in AGB stars: 1) the 
$^{22}$Ne($\alpha$,n)$^{25}$Mg reaction which is activated
at temperatures of $T \gtrsim 300 \times 10^{6}$K, and 2)
the $^{13}$C($\alpha$,n)$^{16}$O reaction, which is 
activated at much lower temperatures of  $T \gtrsim 90 \times
10^{6}$K. Observational and theoretical evidence has shown
that the $^{13}$C($\alpha$,n)$^{16}$O reaction is the
main neutron source in low-mass AGB stars of $\approx 1-3 \textrm{M}_{\odot}$
\citep{straniero95,gallino98,abia02}.

Synthesis by the
s-process in AGB stars is an important contributor to the cosmic abundances past
the iron peak and these stars are also thought to be very important
contributors to the total carbon and nitrogen enrichment \citep[e.g.][]{romano10,kobayashi11}.  Post-AGB
photospheres bear witness to the total chemical changes accumulated
during the stellar lifetime.

In recent years, theoretical models of these internal
nucleosynthesis and photospheric enrichment processes gained
enormously in sophistication: extensive nuclear networks with updated
cross-sections were included \citep[e.g.][]{cristallo11, cristallo09, karakas10a,
  church09}; non-convective mixing such as differential
rotation and thermohaline mixing have been implemented
\citep[e.g.][]{siess03, siess07, stancliffe07, angelou11}; the effect of deep
mixing or extra mixing processes have been critically evaluated
\citep[e.g.][]{karakas10b, busso10} and overshoot regimes have been
explored in more detail in order to explain the extent of the $^{13}$C
pocket \citep{herwig05}.  All these processes involve great
uncertainties and observational data are required to calibrate them.

Accurate determination of photospheric abundances in AGB stars is
difficult \citep[e.g.][and references therein]{abia08}.  First of all,
the photosphere is dominated by molecular opacity, making the mere
detection of trace elements over a wide range of atomic masses
difficult. Moreover, AGB stars often have dynamic atmospheres caused by
pulsations which develop into dust driven winds. Finally, chemical
results of AGB stars are difficult to interpret, because stars with
different initial mass and metallicity occupy the same region in the
HR-diagram.

Post-AGB stars do not have these drawbacks:
First, their atmospheres do not show the large amplitude
pulsations as well as the large mass-loss rates that characterise
evolved AGB atmospheres. Second, their photospheres are hotter, so
atomic transitions prevail. This
allows to quantify the abundances in post-AGB stars for a very wide
range of elements, from CNO up to the most heavy s-process elements,
well beyond the Ba peak \citep[e.g.][]{vanwinckel00, reyniers03}.

During the past decade it has been realised that Galactic post-AGB
stars are chemically much more diverse than anticipated
\citep[e.g.][]{vanwinckel03}.  Some post-AGB stars are indeed the most
s-process enriched objects known to date \citep[e.g.][]{reyniers04},
while others are not enriched at all.  A distinct subclass of Galactic
post-AGB stars is formed by the so-called {\sl 21 $\mu$m objects}
displaying a strong solid-state feature around 21 $\mu$m
\citep{kwok89} in their IR spectra. The carrier of this feature still
needs to be identified, although several suggestions have been
discussed in literature \citep[e.g.][]{posch04}.  The feature is only detected
in post-AGB Carbon stars \citep[e.g.][]{hrivnak08_2,hrivnak09} and till now not yet in spectra of normal carbon
stars nor in carbon-rich planetary nebula \citep{volk11}. The chemical
studies of Galactic 21 $\mu$m stars show that they display strong
overabundances of s-process elements with a wide range of neutron
capture efficiencies \citep{vanwinckel00,reddy02, reyniers03,
  reyniers04}. The poorly known distances and hence luminosities and
masses of the limited Galactic post-AGB sample hamper, however, the
interpretation of the variety of abundances in the broader theoretical
context of stellar (chemical) evolution.

We therefore initiated a project to exploit our newly identified
large sample of post-AGB stars in the Large Magellanic Cloud
\citep{vanaarle11} and in the Small Magellanic Cloud \citep{kamath11}, to
study the s-process production and associated 3rd-dredge-up processes.
The unique spectral characteristics of post-AGB stars, together with
the new large sample, covering a wide range in luminosities and
metallicities, with well constrained distances, means that these
objects provide unprecedented direct tests for the theoretical
structure and enrichment models of solar-mass stars.  In this
contribution, we focus on the object J004441.04-732136.4 (hereafter
abbreviated to J004441.04) which is the only object of the Small
Magellanic Cloud (SMC) known to date in which the 21$\mu$m feature is
identified \citep{volk11}.  The paper is organised as follows: In
Sect. \ref{sect:sample} we discuss the observational data of
J004441.04. Sect. \ref{sect:spec_analysis} describes the spectral
analyses performed to determine the atmospheric parameters as well as
the accurate photospheric composition. The spectral energy
distribution (SED) of J004441.04 is analysed in Sect. \ref{sect:SED}
followed by the determination of the initial mass in Sect.~5.
We then used two state-of-the-art evolution codes coupled to
post-processing codes to compute the predicted chemical signature of
this object in Sect. \ref{sect:models}. The discussion and conclusions
are  given in Sect. \ref{sect:conclusion}.

\section{Data}\label{sect:sample}
\subsection{Photometric data}\label{subsect:photometric_data}

Photometric data from the Magellanic Cloud Photometric Survey (MCPS,
\cite{Zaritsky02}), the Deep Near-Infrared Survey (DENIS,
\cite{Fouque00}), 2MASS \citep{cutri03} and the Spitzer S$^3$MC
survey \citep{bolatto07} are used for the construction of the SED of
J004441.04 in Sect. \ref{sect:SED}. A summary of the absolute
photometric datapoints is given in Table \ref{table:phot}.  The MCPS,
2MASS and DENIS data were taken in a timespan of 2 years (1997 till
1999) while the Spitzer data were acquired in 2004 and 2005. For the
variability analysis, we used the OGLE II lightcurve
\citep{OGLE_05,OGLE_97} obtained in the I filter.

\subsection{Spectroscopic data}\label{subsect:spectroscopic_data}
We use high-resolution spectra obtained with the UVES spectrograph
\citep{dekker00}, which is the echelle spectrograph mounted on the 8m UT2 Kueyen
Telescope of the VLT array at the Paranal Observatory of ESO in
Chili.  Multiple spectra of J004441.04 were obtained on the same day
within a time span of approximately two hours. We used the dichroic
beam-splitter resulting in a wavelength coverage
from approximately 3280 to 4530 \AA{} for
the blue arm of UVES, and from approximately 4780 to 5770 \AA{} and
from 5800 to 6810 \AA{} for the lower and upper part of
the mosaic CCD chip respectively. Each wavelength range was observed three
times with an exposure time of 2871 seconds each.

The UVES reduction pipeline was used for the reduction of the
spectra. This includes the standard steps of extracting frames, determining
wavelength calibration and applying this scale to flat-field divided
data. Also cosmic clipping was included in the reduction.

\begin{figure}[tb!]
\resizebox{\hsize}{!}{\includegraphics{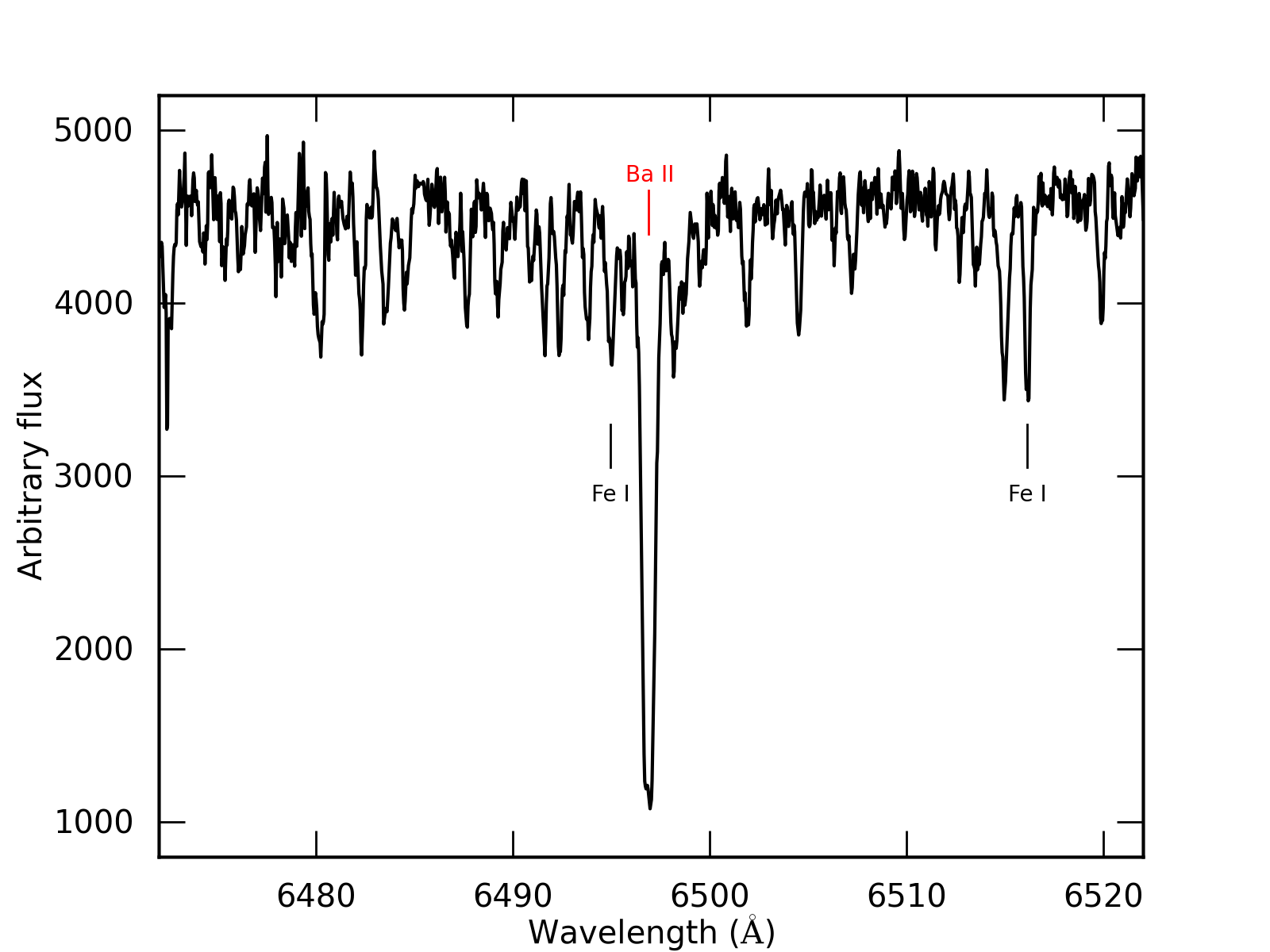}}
\caption{Not-normalised UVES spectrum of J004441.04, centered around
  the Ba II line at 6496.897 \AA{}. The spectrum is set to rest velocity.}\label{fig:ori}
\end{figure}

All spectra were normalised by fitting in small spectral windows, fifth order
polynomials through interactively defined
continuum points.
 Once all three subspectra of the same wavelength
range are normalised, the weighted mean subspectrum was determined. These mean subspectra are then again merged
into one normalised spectrum and this is used for the spectral abundance
determination.

Unfortunately, the S/N is too poor for a large part of the blue
spectrum (wavelength range 3280 to 4200 \AA{}) making these wavelength
ranges unusable for an accurate spectral abundance analysis hence they
are not used for the study of J004441.04. Except for these bluer
wavelength ranges, an overall S/N around 100 is obtained.

\begin{table}[t]
\caption{\label{table:phot} Photometric data used for the SED of J004441.04. A standard error of 0.05 mag is assumed for the SED study. IRAC and MIPS are instruments of the Spitzer satellite.}
\begin{center}
\begin{tabular}{lccc} \hline\hline
band  & survey & mag & $\textrm{F}_{\nu}$ (Jy) \\
\hline
V               & MCPS    & 15.96  & 1.573 $\times 10^{-3}$  \\
I               & MCPS    & 14.46  & 5.635 $\times 10^{-3}$  \\
$\textrm{I}_c$  & DENIS   & 14.44  & 4.260 $\times 10^{-3}$  \\
J               & DENIS   & 13.62  & 6.310 $\times 10^{-3}$  \\
K               & DENIS   & 12.94  & 4.194 $\times 10^{-3}$  \\
J               & 2MASS   & 13.63  & 5.635 $\times 10^{-3}$  \\
H               & 2MASS   & 13.23  & 5.216 $\times 10^{-3}$  \\
$\textrm{K}_S$  & 2MASS   & 13.02  & 4.128 $\times 10^{-3}$  \\
$[3.6]$           & IRAC    & 12.33  & 3.976 $\times 10^{-3}$  \\
$[4.5]$           & IRAC    & 11.90  & 3.434 $\times 10^{-3}$  \\
$[5.8]$           & IRAC    & 10.39  & 1.010 $\times 10^{-2}$  \\
$[8.0]$           & IRAC    & 8.344  & 3.459 $\times 10^{-2}$  \\
$[24]$            & MIPS    & 5.141  & 5.998 $\times 10^{-2}$  \\
\hline
\end{tabular}
\end{center}
\end{table}

\section{Spectral analyses}\label{sect:spec_analysis}

\begin{figure}
\resizebox{\hsize}{!}{\includegraphics{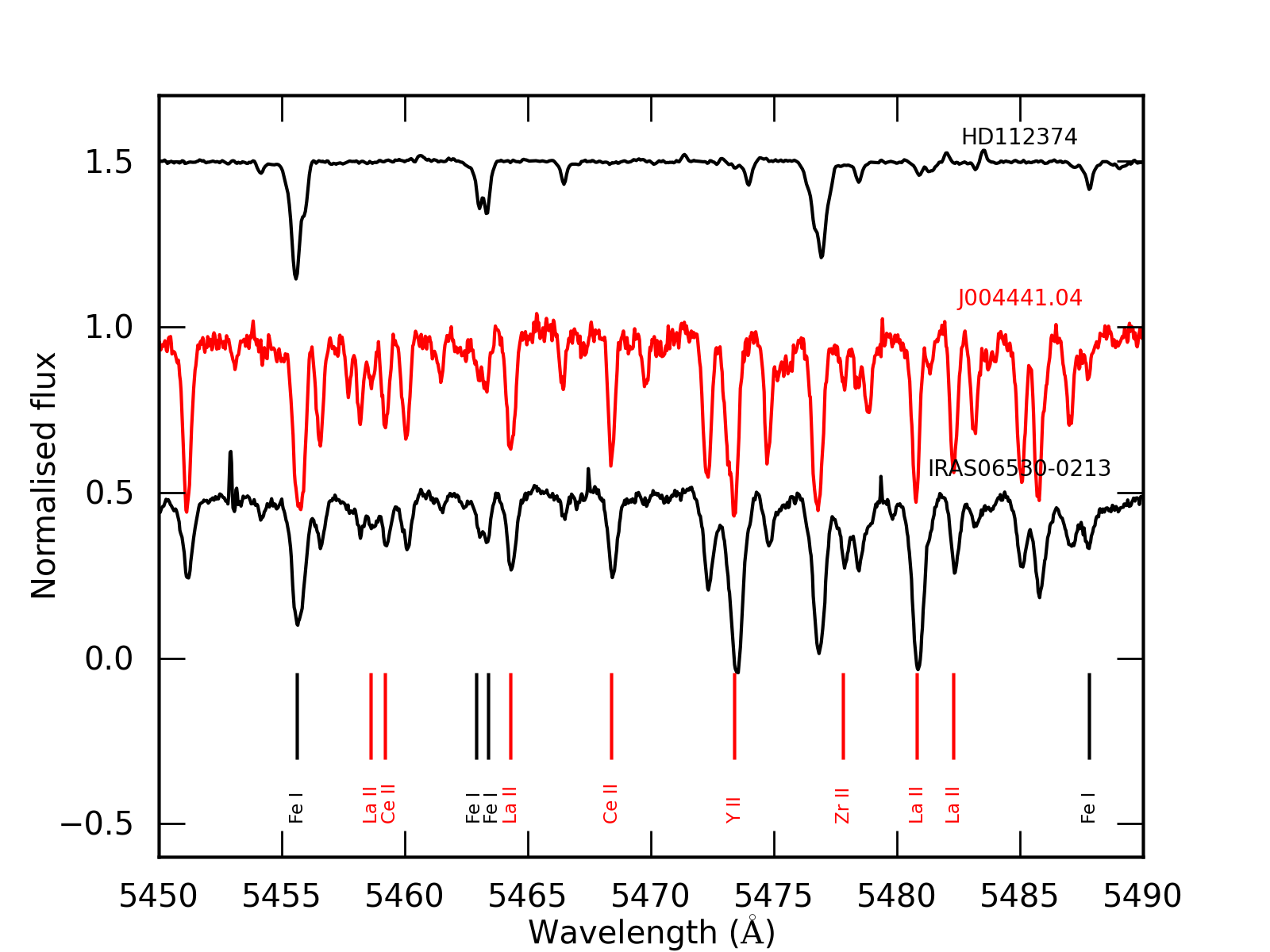}}
\caption{Comparison of the normalised spectra of J004441.04 (middle), IRAS06530 (lower) and HD112374 (upper). The upper and lower spectra have been shifted in flux for clarity. Red and black vertical lines mark positions of s-nuclei and non s-nuclei respectively. For more information, see text.}\label{fig:comp_J0044}
\end{figure}

\begin{figure}
\resizebox{\hsize}{!}{\includegraphics{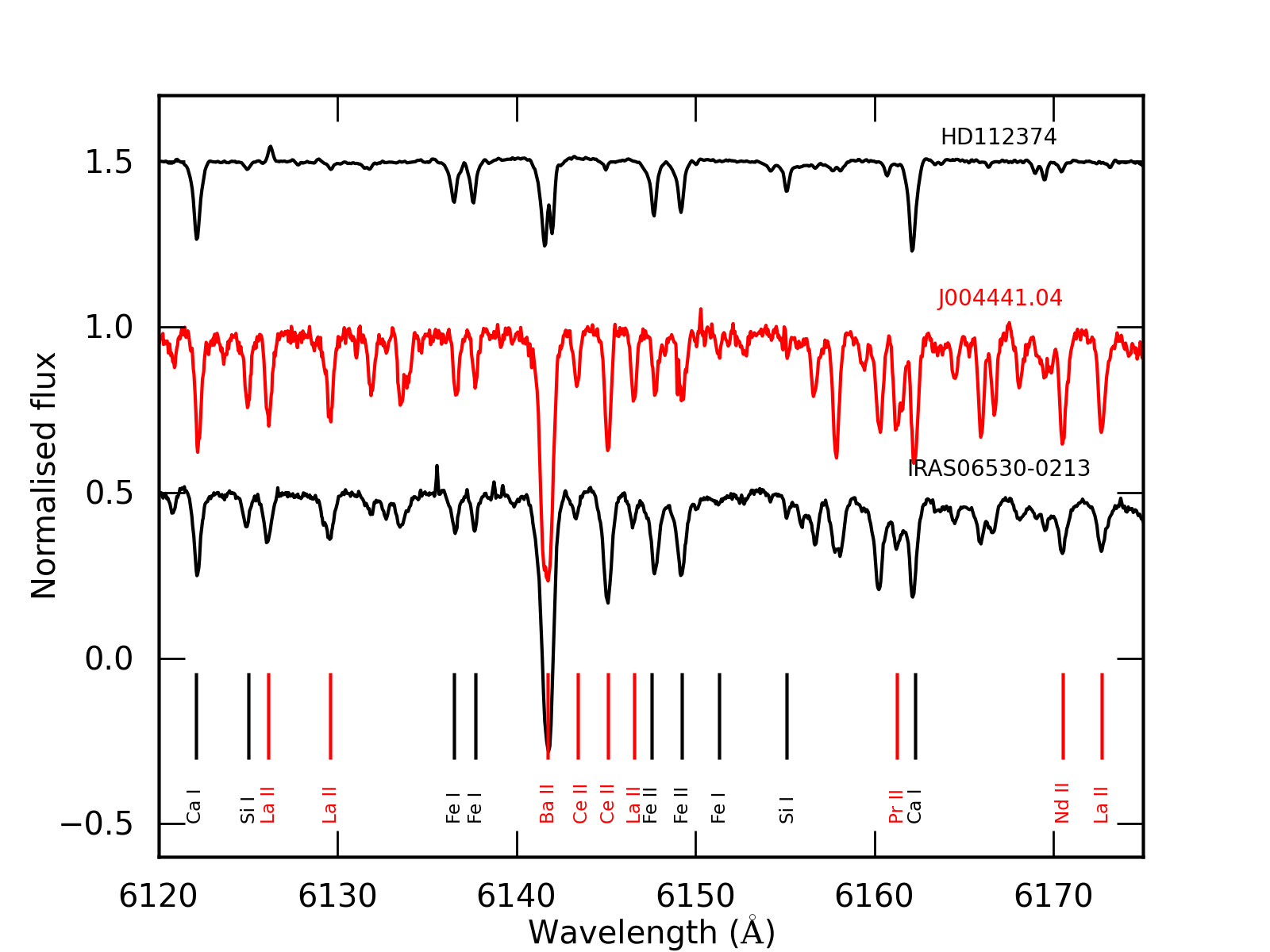}}
\caption{Comparison of the normalised spectra of J004441.04 (middle), IRAS06530 (lower) and HD112374 (upper). The upper and lower spectra have been shifted in flux for clarity. Red and black vertical lines mark positions of s-nuclei and non s-nuclei respectively.}\label{fig:vb}
\end{figure}

It is useful to compare the spectrum of J004441.04 with Galactic
post-AGB stars.
In Fig. \ref{fig:comp_J0044} and \ref{fig:vb} we compare a small
spectral part of J004441.04 with the same spectral range of two
Galactic post-AGB stars. The
lower spectrum is from the Galactic post-AGB star IRAS06530-0213 (hereafter abbreviated to
IRAS06530). This is a post-AGB star with a metallicity of $[\textrm{Fe/H}]$ =
-0.46 and $\textrm{T}_{\rm eff}$ = 7250 K and it is recognised to be the star
with the highest s-process overabundances known to date  \citep{reyniers04}. 
The upper spectrum is from
the non-enriched Galactic star HD112374 = HR 4912
\citep{lambert83,giridhar97}. This star has a similar spectral type
and a metallicity of [Fe/H]=$-$1.1.
The normalised spectra  are brought to the same
velocity scale and offset for clarity. Red and black vertical lines
mark positions of s-nuclei and non s-nuclei respectively. 
It is remarkable that the spectra of J004441.04 are very similar to 
IRAS06530-0213 which illustrates its high s-process overabundance. 
We therefore choose the used line list of
IRAS06530 in \citet{reyniers04} for quantified relative
spectral analyses.


\subsection{General methods}\label{subsect:methods}
Both the atmospheric parameter determination and the abundance
determination are performed using the same routines. The local thermal
equilibrium (LTE) Kurucz-Castelli atmosphere models \citep{castelli04}
are used in combination with the LTE abundance calculation routine
MOOG (version July 2009) by C. \citet{sneden73}. Non-LTE effects are
not taken into account for the spectral analyses of
J004441.04. 

The equivalent width (EW) of lines are measured via direct
integration and abundances are computed by an iterative process in which
the theoretical EWs of single lines are computed for a given abundance
and matched to the observed EWs. Blended lines are avoided as much as
possible. To check whether lines are part of blends with other
identified lines, synthetic spectra are modelled with MOOG using VALD
linelists \citep{vald} and compared to the stellar spectra.

In order to identify lines, the radial velocity of
J004441.04 is estimated by fitting a Gaussian curve to a number of
identified atomic lines to determine their central wavelength. Using
the equation of the Doppler shift, this results in a heliocentric radial velocity v
= 148 $\pm$ 3 km/s which is accurate enough for line identification
purposes. The useful line identification tool of \cite{lobel06} was
used for the first identifications. The heliocentric radial velocity of the SMC is in average
160 km/s \citep{richter87} and the velocity of J004441.04 confirms the
membership of the SMC.

\subsection{Atmospheric parameter determination}\label{subsect:atd}

The atmospheric parameters are derived using Fe I and Fe II lines and
the standard spectroscopic methods: the effective temperature
$\textrm{T}_{\rm eff}$ is derived by imposing the iron abundance, derived
from individual Fe I lines, to be independent of lower excitation
potential; surface gravity log g is derived by imposing ionization
equilibrium between the iron abundance of individual Fe I and Fe II;
the microturbulent velocity $\xi_t$ is derived by imposing the iron
abundance from individual Fe I lines to be independent of reduced
equivalent width (RW). For these analysis, 17 Fe I and 4 Fe II lines
are used. Due to the high s-process enrichment of J004441.04, the
number of useful non-blended Fe lines for the atmospheric determination is
limited, especially for Fe II.
\begin{figure}[tb!]
\resizebox{\hsize}{!}{\includegraphics{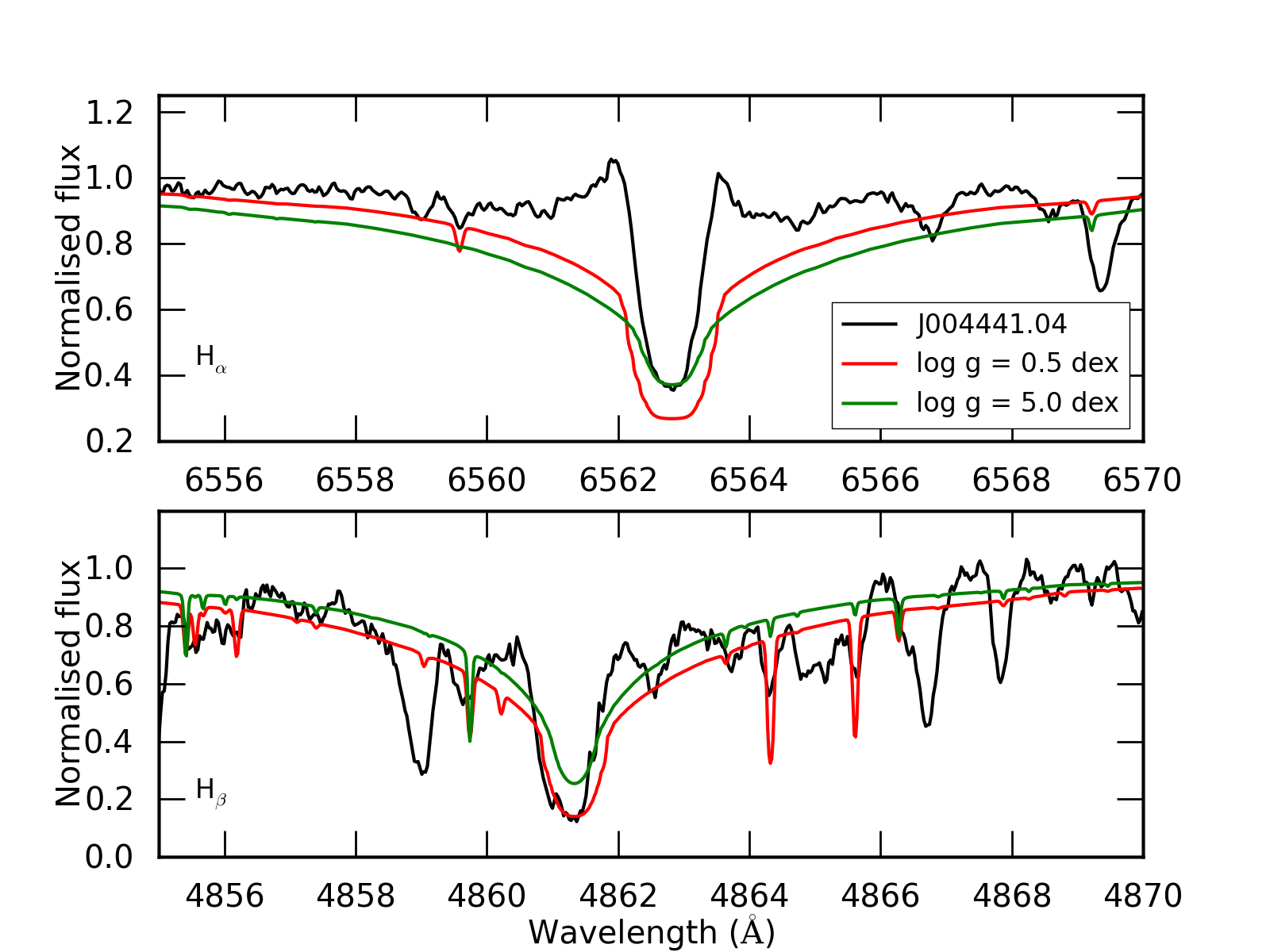}}
\caption{Comparison between Balmer lines of J004441.04 with synthetic models of \citet{coelho05} with different log g. The upper panel shows $\textrm{H}_{\alpha}$, the lower panel shows $\textrm{H}_{\beta}$. The black spectrum represents J004441.04, the red and green spectrum are synthetic spectra with a log g of 0.5 and 5.0 dex respectively. For more information, see text.}\label{fig:H_comp}
\end{figure}

The results for the different atmospheric parameters are shown in
Table \ref{table:atmos} and lie in the range of typical post-AGB
parameter values. The iron abundance for the determined atmospheric
parameters is a good estimate of the overall chemical composition
$\left[\textrm{M/H}\right]$ which is in this case
$\left[\textrm{Fe/H}\right]$ = -1.34 $\pm$ 0.32 (dex). The abundance
error of $\pm$ 0.32 is the total error including the line to line
scatter and the uncertainties by the used atmospheric model which will
be further discussed in Sect. \ref{subsect:abun}. The low iron
abundance of J004441.04 with respect to the mean metallicity of the
SMC of $\left[\textrm{Fe/H}\right] \approx -0.7$ \citep{luck98}
classifies J004441.04 as a low-metallicity star which are generally
acknowledged as astrophysical production sites of heavy s-nuclei
provided the third dredge-up (TDU) takes place.
 
\begin{table}[tb!]
\caption{\label{table:atmos} Determined atmospheric parameters of J004441.04}
\begin{center}
\begin{tabular}{lc} \hline
$\textrm{T}_{\rm eff}$ (K) & $6250 \pm 250$ \\
log g (dex) & $0.5 \pm 0.5$ \\
$\xi_t$ (km/s) & $3.5 \pm 0.5$ \\
$\left[\textrm{M/H}\right]$ (dex) & -1.34 $\pm$ 0.32 \\
\hline
\end{tabular}
\end{center}
\end{table}

An attempt was made to determine the surface gravity of
J004441.04 by fitting synthetic spectra to its Balmer lines
$\textrm{H}_{\alpha}$ and $\textrm{H}_{\beta}$ at 6562.8 \AA{} and
4861.3 \AA{} respectively. The comparison between the Balmer wings of
the synthetic models with different log g and J004441.04 would provide
an estimate of the surface gravity of J004441.04. However, this method
is not usable for J004441.04, since its Balmer wings display emission
lines as shown in Fig. \ref{fig:H_comp}. The upper panel shows the
comparison between $\textrm{H}_{\alpha}$ lines, the lower panel shows
$\textrm{H}_{\beta}$. The black spectrum represents J004441.04, the
red and green spectrum are synthetic spectra of \citet{coelho05} with
a log g of 0.5 and 5.0 dex respectively. Both panels display emission
in the wings of both Balmer lines of J004441.04 which are most
probably caused by ongoing mass loss. Also the strong enrichment of J004441.04
makes it impossible to fit the Balmer wings, since J004441.04 displays
multiple strong atomic lines which are not included in the synthetic
spectra. This is especially clear in the lower panel of
Fig. \ref{fig:H_comp}.

The atmospheric parameters can be checked using results on other
species only when the number of lines for that species is
significant. 
Unfortunately, most species only have two, three or four
useful lines (see already Table \ref{table:abun}) due to the high
number of blends.  Fe I is the only neutral ion with single lines covering a large
range in excitation potential. Fe is also the only element for which a
useful number of lines of different ions is found. Therefore, the microturbulent velocity $\xi_t$ is the
only parameter which can be checked using another species than Fe:
although the used number of lines of \ion{La}{II}, \ion{Ce}{II} and \ion{Nd}{II}
is small, they do provide a mean to check the derived $\xi_t$.
This $\xi_t$ check with \ion{La}{II} and \ion{Ce}{II} yields
$\xi_t = 3.5$ km/s and \ion{Nd}{II} gives $\xi_t = 3.0$ km/s
, which confirms the found microturbulent velocity.

\subsection{Abundance determination}\label{subsect:abun}

With our preferred model atmosphere as basis, we started with a full
but strictly relative abundance analysis and limited
ourselves first to the lines used in the spectral analyses of IRAS06530-0213.

We used mainly isolated non-blended, non-saturated lines and
the individual atomic lines are all double-checked via a spectrum
synthesis to investigate the possible presence of unresolved blends.
N lines with EWs
larger than 3 \AA{} are not found and are not used due to possible
confusion with noise in the spectrum. A redder
spectrum will be needed for the N abundance determination.

\begin{figure}[tb!]
\begin{center}
\includegraphics[width=9cm]{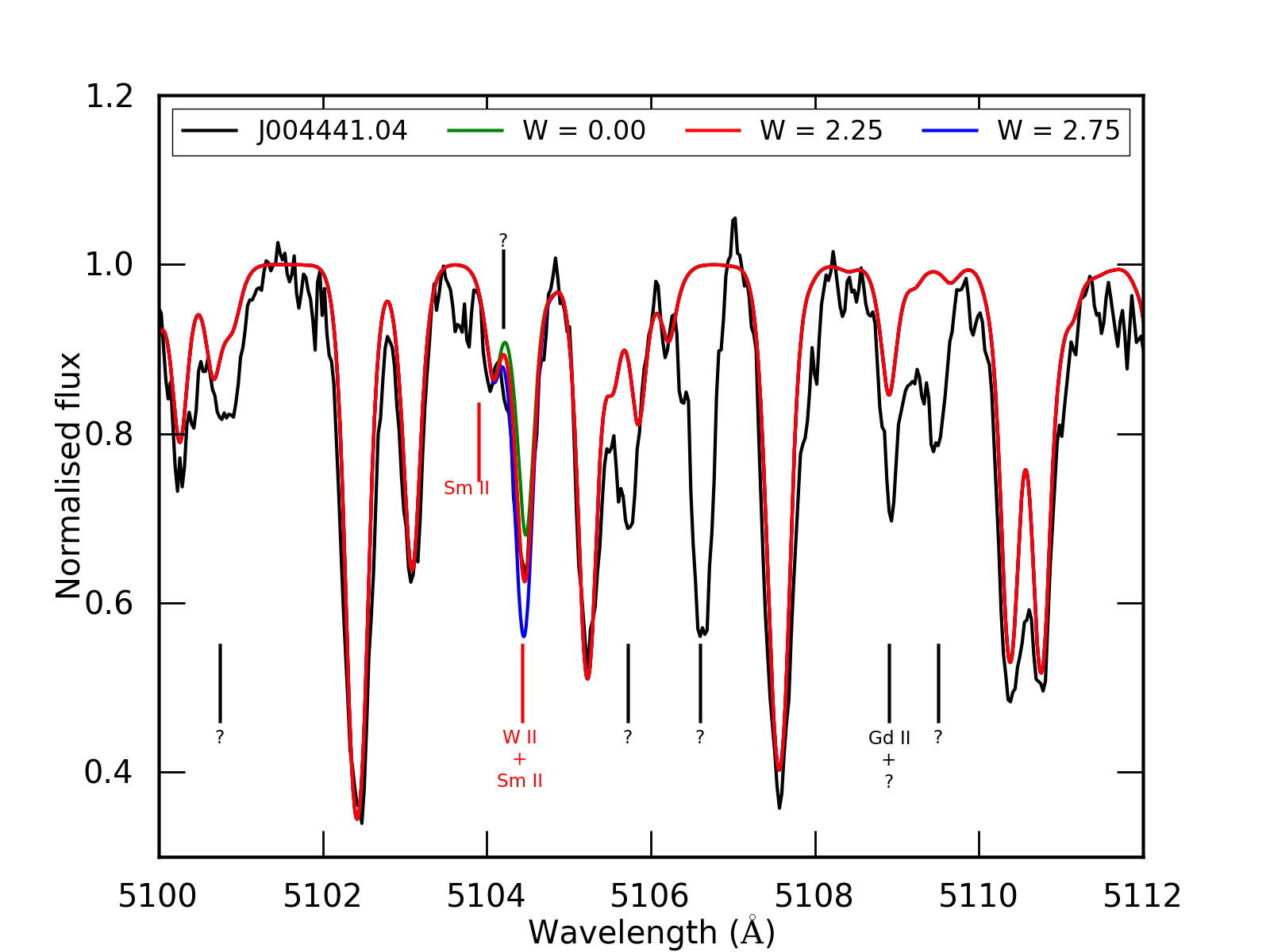}
\end{center}
\caption{Spectrum synthesis of the W II line at 5104.432 \AA{}. The black spectrum is J004441.04, the colored spectra are synthetic spectra with different W abundances. The Sm II line at 5104.080 \AA{}, which forms a blend with W II, is indicated together with some atomic lines which are unknown in the line list. For more information, see text.}\label{fig:74WII}
\end{figure}

\begin{figure}[tb!]
\begin{center}
\includegraphics[width=9cm]{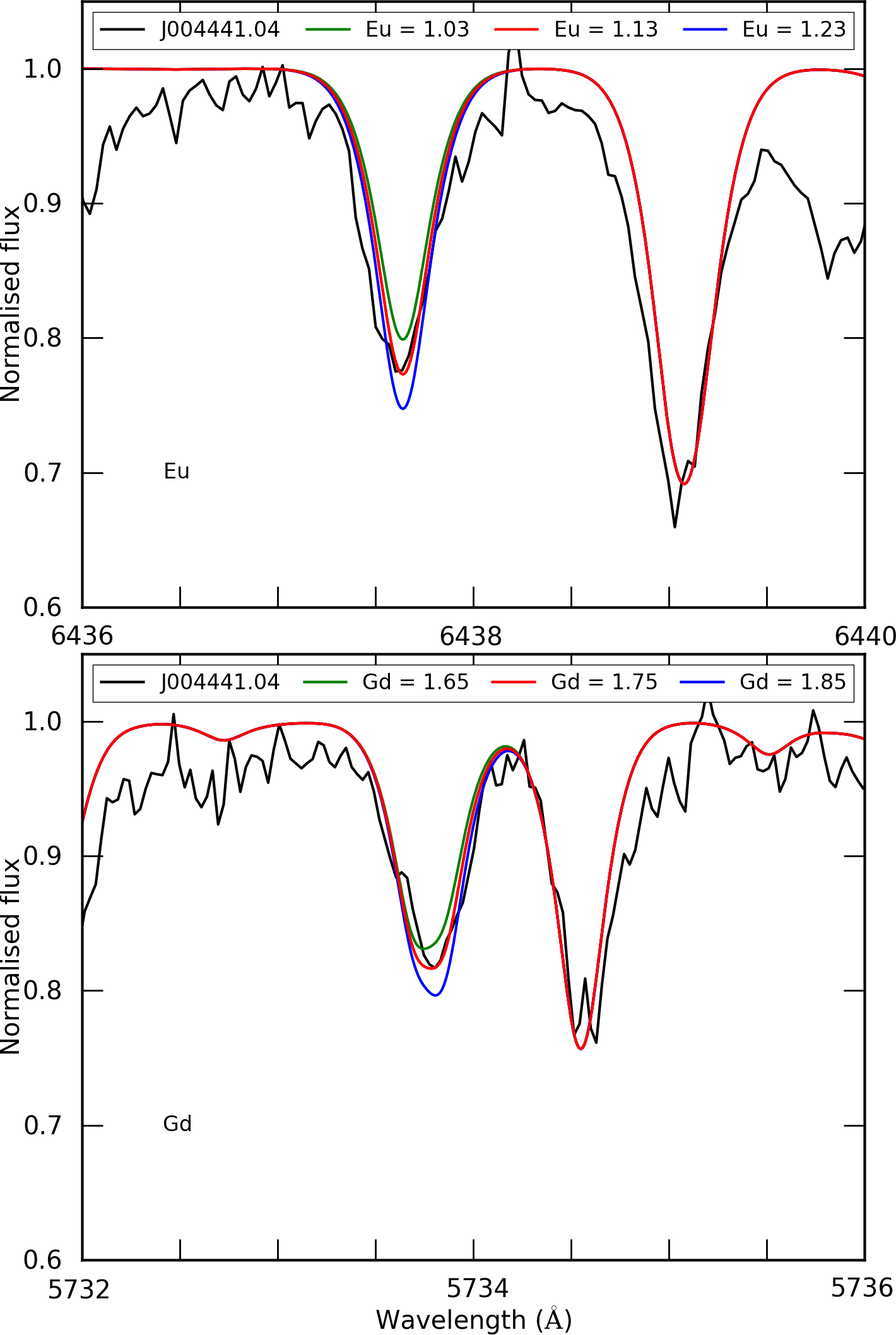}
\caption{Spectrum synthesis of Eu II (top) and Gd II (bottom) lines at 6437.640 \AA{} and 5733.852 \AA{} respectively. The black spectrum is J004441.04, the colored spectra are synthetic spectra with different abundances of the studied element. For more information, see text.}\label{fig:synthesis}
\end{center}
\end{figure}

Unfortunately, at these high
overabundances, all Sr and Ba lines are heavily saturated making
accurate abundance determination of these two most famous s-process
species impossible (Figs. \ref{fig:ori} and  \ref{fig:vb}). The other
s-process abundances come from isolated single lines except for those
species where all detectable lines turned out to be blended.
These specific blends were fitted by
creating synthetic spectra in MOOG using VALD line lists of specific
wavelength ranges. The resulting fits for Eu II and Gd II are shown in
Fig. \ref{fig:synthesis} for the Eu II line at 6437.640 \AA{} in the
upper panel and the Gd II line at 5733.852 \AA{} in the lower
panel. The black spectrum is the spectrum of J004441.04, the colored
spectra are synthetic spectra with different abundances of the studied
element. The Eu II line at 6437.640 \AA{} is part of a blend with a
very weak Y II line at 6437.169 \AA{}, the Gd II line is part of a
blend with a relatively strong Ce II line at 5733.692 \AA{}. 
Fig. \ref{fig:74WII} shows the spectrum synthesis of the W II
line at 5104.432 \AA{}. The black spectrum is J004441.04, the colored
spectra are synthetic spectra with different W abundances. This W II
line forms a blend with a Sm II line at 5104.479 \AA{} which is
indicated in the figure. The positions of lines unknown to VALD are
indicated with '$?$' proving that there are still a large number of
lines that need to be identified in these highly enriched object. 

The next step in the process was to constrain abundances for elements
which have no isolated single lines. 
We used VALD line lists to predict the equivalenth widths for all
lines of a given element by assuming a strong overabundance.
For each element we then choose the strongest line in our observed spectrum and 
fitted this line with synthetic spectra of MOOG, analog to the Eu II and Gd II 
abundances so with appropriate input abundances of all elements which
we could quantify previously. Unfortunately, also here the strong overabundances of s-process 
elements create spectral lines which are unknown to VALD linelists making 
abundance determinations difficult because of possible blends of
unknown origin. Using the 
synthetic spectrum method, we estimated abundances for Mg I and Zn I together 
with the s-process elements Dy II, Er II, Yb II, Lu II and Hf II. Afterwards, 
we determined the EW of each used line in order to perform an error analysis 
on the derived abundances.
For the abundance determination of O, we did not use the forbidden O I lines at 6300 
and 6363 \AA{} as they are part of an identified 
and unidentified blend respectively. After the abundance determination, we performed a spectral 
synthesis of the 6300 \AA{} line using the derived O abundance which rendered a good fit. As 
forbidden lines are not sensitive to non-LTE effects (see
e.g. \cite{kiselman02}), this is a confirmation of the obtained
high oxygen abudance.

The final abundance results of J004441.04 are shown in Table
\ref{table:abun} where N represents the number of lines used for the
abundance determination of the species, the full line list used for
the abundance determination is shown in Table
\ref{tab:all_lines}. Solar abundances are taken from
\citet{asplund09}, uncertanties in log $\epsilon$ and log
$\epsilon_{\sun}$ due to line-to-line scatter are given by
$\sigma_{\epsilon}$ and $\sigma_{\epsilon_{\sun}}$ respectively. The
sensitivity of $[\textrm{X/Fe}]$ to uncertainties in the line-to-line
scatter, effective temperature, surface gravity and microturbulent
velocity are given by $\sigma_{[X/Fe]}$, $\sigma_{T_{eff}}$,
$\sigma_{log g}$ and $\sigma_{\xi_{t}}$ respectively. The combined
uncertainty in $[\textrm{X/Fe}]$ due to the above described
sensitivity is represented by $\sigma_{tot}$. Elemental abundances derived via synthetic 
spectrum fitting are indicated in an italic font in Table \ref{table:abun}. We remark again that
non-LTE effects are not taken into account in the abundance
determination of the different species. 
Although the abundances of some useful elements for nucleosynthesis
studies cannot be determined, Table \ref{table:abun} still contains
quantified abundances of a wide range of s-elements.

\begin{table*}[t!]
\caption{\label{table:abun} Abundance results of J004441.04 together with the calcuted errors due to line to line scatter and uncertainties of the chosen atmosphere model. N represents the number of lines used for the abundance determination of the species. Uncertainties in log $\epsilon$ and log $\epsilon_{\sun}$ due to line-to-line scatter are given by $\sigma_{\epsilon}$ and $\sigma_{\epsilon_{\sun}}$ respectively. $\sigma_{[X/Fe]}$, $\sigma_{T_{eff}}$, $\sigma_{log g}$ and $\sigma_{\xi_{t}}$ represent the sensitivity of $[\textrm{X/Fe}]$ to uncertainties in the line-to-line scatter, effective temperature, surface gravity and microturbulent velocity respectively. $\sigma_{tot}$ represents the total uncertainty of $[\textrm{X/Fe}]$ due to the line-to-line scatter and model uncertainties. Solar abundances are retrieved from \citet{asplund09}. Abundances of elements indicated with italicised text are determined via synthetic spectra fitting. For more information, see text.}
\begin{center}
\begin{tabular}{lccccccccccc} \hline\hline
Species & N & log $\epsilon \pm \sigma_{\epsilon}$ &  log $\epsilon_{\sun} \pm \sigma_{\epsilon_{\sun}}$ & $[\textrm{X/Fe}] \pm \sigma_{tot}$ & $\sigma_{[X/Fe]}$ & \multicolumn{2}{c}{$\sigma_{T_{eff}}$} & $\sigma_{log g}$ & \multicolumn{2}{c}{$\sigma_{\xi_{t}}$} \\
 & & & & & & +250 K & -250 K & +0.5 dex& +0.5 km/s & -0.5 km/s \\
\hline
C I & 6 & $8.76 \pm 0.10$ & $8.43 \pm 0.05 $ & $ 1.67 \pm 0.36 $  & 0.14 & -0.23 & 0.24 & 0.11 & 0.02 & -0.02\\
O I & 3 & $8.49 \pm 0.09$ & $8.69 \pm 0.05 $ & $ 1.14 \pm 0.50 $  & 0.13 & -0.35 & 0.32 & 0.12 & 0.02 & -0.05\\
Na I & 2 & $5.85 \pm 0.08$ & $6.24 \pm 0.04 $ & $ 0.95 \pm 0.14 $  & 0.12 & -0.08 & 0.06 & -0.02 & 0.01 & -0.03\\
\textit{Mg I} & \textit{1} & $\mathit{6.75 \pm 0.20}$ & $\mathit{7.60 \pm 0.04 }$ & $ \mathit{0.49 \pm 0.26 }$  & \textit{0.22} & \textit{-0.11} & \textit{0.05} & \textit{-0.05} & \textit{0.02} & \textit{-0.04}\\
Si I & 2 & $6.63 \pm 0.01$ & $7.51 \pm 0.03 $ & $ 0.46 \pm 0.16 $  & 0.09 & -0.11 & 0.07 & -0.05 & 0.02 & -0.04\\
S I & 2 & $6.07 \pm 0.13$ & $7.12 \pm 0.03 $ & $ 0.29 \pm 0.26 $  & 0.16 & -0.17 & 0.15 & 0.03 & 0.02 & -0.04\\
Ca I & 4 & $5.33 \pm 0.03$ & $6.34 \pm 0.04 $ & $ 0.33 \pm 0.06 $  & 0.09 & -0.03 & 0.03 & 0.01 & 0.00 & 0.00\\
Sc II & 3 & $2.18 \pm 0.08$ & $3.15 \pm 0.04 $ & $ 0.35 \pm 0.14 $  & 0.15 & 0.08 & -0.08 & 0.01 & -0.01 & 0.01\\
Ti II & 3 & $4.02 \pm 0.21$ & $4.95 \pm 0.05 $ & $ 0.39 \pm 0.16 $  & 0.25 & 0.06 & -0.05 & 0.01 & -0.02 & 0.01\\
Cr II & 3 & $4.26 \pm 0.04$ & $5.64 \pm 0.04 $ & $ -0.06 \pm 0.08 $  & 0.13 & -0.02 & 0.00 & -0.02 & 0.01 & -0.02\\
Mn I & 2 & $4.77 \pm 0.11$ & $5.43 \pm 0.04 $ & $ 0.68 \pm 0.14 $  & 0.14 & -0.07 & 0.01 & -0.05 & 0.01 & -0.04\\
Fe I & 17 & $6.16 \pm 0.07$ & $7.50 \pm 0.04 $ & $ 0.00 \pm 0.03 $  & 0.11 & 0.00 & 0.00 & 0.00 & 0.00 & 0.00\\
Fe II & 4 & $6.18 \pm 0.11$ & $7.50 \pm 0.04 $ & $ 0.00 \pm 0.08 $  & 0.17 & 0.00 & 0.00 & 0.00 & 0.00 & 0.00\\
Ni I & 4 & $5.09 \pm 0.17$ & $6.22 \pm 0.04 $ & $ 0.21 \pm 0.13 $  & 0.19 & -0.05 & 0.01 & -0.04 & 0.03 & -0.04\\
\textit{Zn I} & \textit{1} & $\mathit{3.85 \pm 0.20}$ & $\mathit{4.56 \pm 0.05} $ & $\mathit{ 0.63 \pm 0.23} $  & \textit{0.22} & \textit{-0.06} & \textit{0.00} & \textit{-0.04} & \textit{-0.01} & \textit{0.00}\\
Y II & 2 & $3.04 \pm 0.03$ & $2.21 \pm 0.05 $ & $ 2.15 \pm 0.42 $  & 0.13 & 0.26 & -0.08 & 0.20 & -0.13 & 0.19\\
Zr II & 2 & $3.23 \pm 0.11$ & $2.58 \pm 0.04 $ & $ 1.97 \pm 0.22 $  & 0.17 & 0.13 & -0.08 & 0.07 & -0.06 & 0.07\\
La II & 9 & $2.62 \pm 0.13$ & $1.10 \pm 0.04 $ & $ 2.84 \pm 0.32 $  & 0.18 & 0.23 & -0.14 & 0.09 & -0.08 & 0.10\\
Ce II & 8 & $2.79 \pm 0.13$ & $1.58 \pm 0.04 $ & $ 2.53 \pm 0.18 $  & 0.18 & 0.12 & -0.12 & -0.01 & -0.02 & 0.01\\
Pr II & 2 & $2.10 \pm 0.01$ & $0.72 \pm 0.04 $ & $ 2.70 \pm 0.32 $  & 0.12 & 0.23 & -0.18 & 0.02 & -0.06 & 0.06\\
Nd II & 7 & $2.84 \pm 0.10$ & $1.42 \pm 0.04 $ & $ 2.74 \pm 0.31 $  & 0.16 & 0.24 & -0.17 & 0.03 & -0.05 & 0.07\\
Sm II & 2 & $1.85 \pm 0.04$ & $0.96 \pm 0.04 $ & $ 2.21 \pm 0.30 $  & 0.13 & 0.22 & -0.16 & 0.05 & -0.05 & 0.07\\
\textit{Eu II} & \textit{1} & $\mathit{1.13 \pm 0.10}$ & $\mathit{0.52 \pm 0.04 }$ & $\mathit{ 1.93 \pm 0.24 }$  &\textit{ 0.16} & \textit{0.10} & \textit{-0.13} & \textit{-0.06} & \textit{-0.04} & \textit{0.04}\\
 \textit{Gd II} & \textit{1} & $\mathit{1.75 \pm 0.10}$ & $\mathit{1.07 \pm 0.04} $ & $\mathit{ 2.00 \pm 0.20 }$  &\textit{ 0.16} & \textit{0.05} & \textit{-0.10} & \textit{-0.04} & \textit{0.00} & \textit{-0.02}\\
\textit{Dy II} & \textit{1} & $\mathit{2.05 \pm 0.20}$ & $\mathit{1.10 \pm 0.04 }$ & $\mathit{ 2.27 \pm 0.22 }$  &\textit{ 0.12} & \textit{0.11} & \textit{-0.16} & \textit{-0.05} & \textit{-0.01} & \textit{-0.01}\\
\textit{Er II} & \textit{1} & $\mathit{1.90 \pm 0.20}$ & $\mathit{0.92 \pm 0.05 }$ & $\mathit{ 2.30 \pm 0.25 }$  &\textit{ 0.24} & \textit{0.04} & \textit{-0.07} & \textit{-0.04} & \textit{0.01} & \textit{-0.02}\\
\textit{Yb II} & \textit{1} & $\mathit{2.15 \pm 0.20}$ & $\mathit{0.84 \pm 0.11 }$ & $ \mathit{2.63 \pm 0.63 }$  &\textit{ 0.26} & \textit{-0.02} & \textit{-0.01} & \textit{-0.04} & \textit{0.57} & \textit{0.00}\\
\textit{Lu II} & \textit{1} & $\mathit{1.30 \pm 0.20}$ & $\mathit{0.10 \pm 0.09 }$ & $ \mathit{2.52 \pm 0.15 }$  &\textit{ 0.15} & \textit{0.02} & \textit{-0.06} & \textit{-0.03} & \textit{-0.04} & \textit{0.06}\\
\textit{Hf II} & \textit{1} & $\mathit{2.15 \pm 0.20}$ & $\mathit{0.85 \pm 0.04 }$ & $\mathit{ 2.62 \pm 0.13 }$  &\textit{ 0.12} & \textit{0.03} & \textit{-0.08} & \textit{-0.03} & \textit{0.00} & \textit{-0.01}\\
\textit{W II} & \textit{1} & $\mathit{2.25 \pm 0.20}$ & $\mathit{0.85 \pm 0.12 }$ & $\mathit{ 2.72 \pm 0.27 }$  &
\textit{0.26} & \textit{0.02} & \textit{-0.07} & \textit{-0.04} & \textit{-0.01} & \textit{-0.01}\\
\hline
\end{tabular}
\end{center}
\end{table*}

\subsection{Abundance results}\label{subsect:abunresults}
\begin{figure}[tb!]
\resizebox{\hsize}{!}{\includegraphics{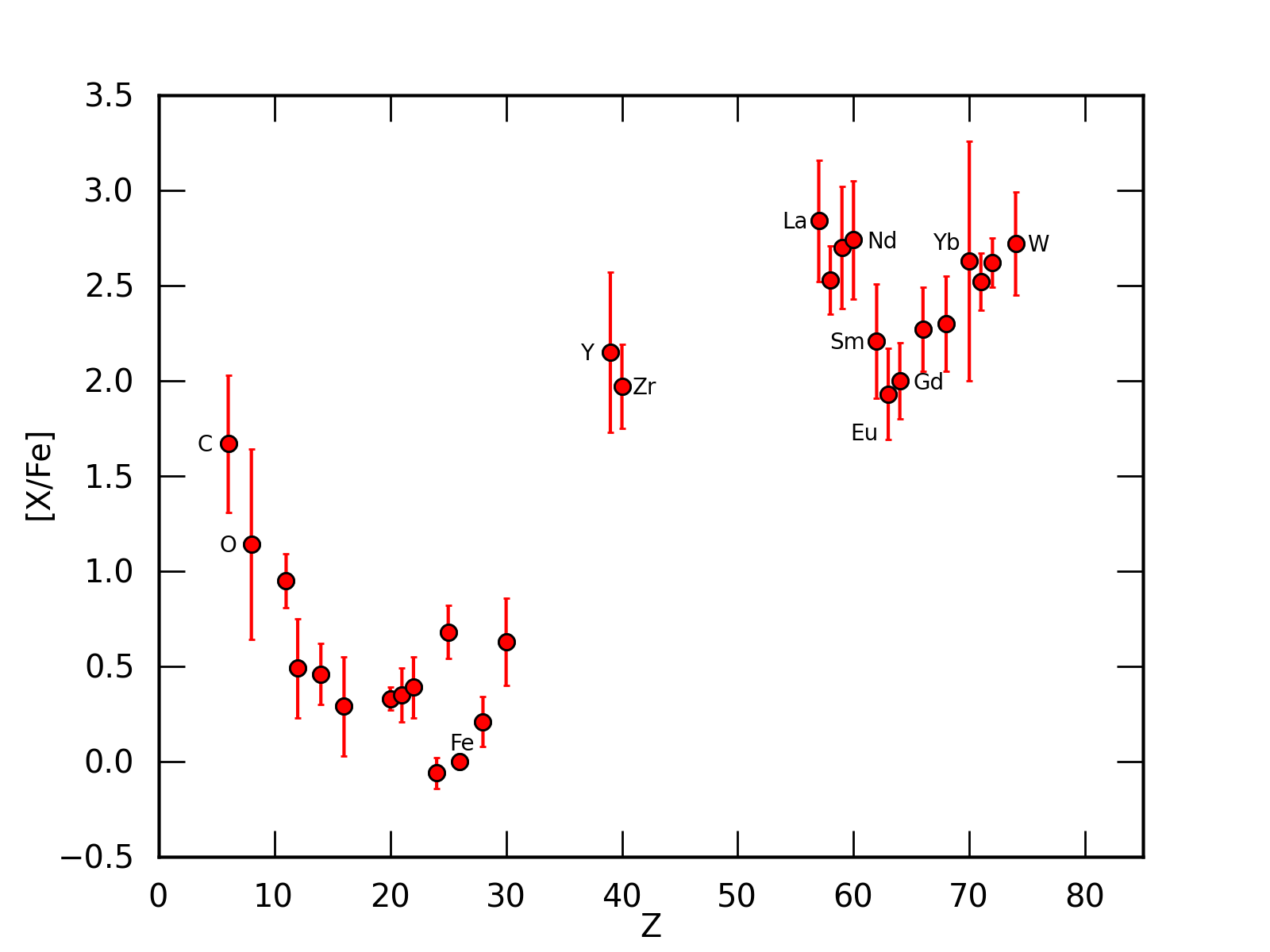}}
\caption{$[\textrm{X}/\textrm{Fe}]$ results of J004441.04, the errorbars represent the total uncertainty $\sigma_{tot}$. Some elements are labelled for clarity. For more information, see text.}\label{fig:XFe}
\end{figure}

The $[\textrm{X/Fe}]$ results of Table \ref{table:abun} are plotted in
Fig. \ref{fig:XFe}, the errorbars represent the total uncertainty
$\sigma_{tot}$. For clarity some elements are labelled. The figure
illustrates well the post-carbon star signature of J004441.04, with a C enrichement
and very strong s-process enrichment. The C/O ratio is 1.9 $\pm$ 0.7.
The uncertainties of the model atmosphere has the biggest impact on
the accuracy of this ratio. J004441.04 is a post-AGB Carbon star with a C/O larger
than one, but only mildly so.

Concerning the available $\alpha$-elements Mg, Si, S, Ca and Ti, the
simple mean of the $[\textrm{X/Fe}]$ is $[\alpha/\textrm{Fe}] =
+0.4$. Such an enhancement is normal for Galactic objects in this
metallicity range as a consequence of galactic chemical evolution and
would not point to an intrinsic enhancement. However, it is not clear
whether this relation also holds for the SMC.

The extreme s-process abundances point to
a very effective dredge-up. With [Y/Fe] = +2.2 and [La,Ce,Pr,Nd/Fe]
all inbetween 2.5 and 2.8 the s-process overabundance is extreme.
Also the s-process overabundances of elements well beyond the Ba peak are high. 
The abundance results show that J004441.04 is the strongest s-process enriched extra-galactic
object studied to date.

Fig. \ref{fig:XFe} shows a stronger
overabundance of the heavy s-process component (magic neutron number
82) around La and Ce with respect to the light s-process component
(magic neutron number 50) around Y and Zr which indicates an effective
$^{13}$C pocket for the production of heavy elements. 

\begin{figure}
\resizebox{\hsize}{!}{\includegraphics{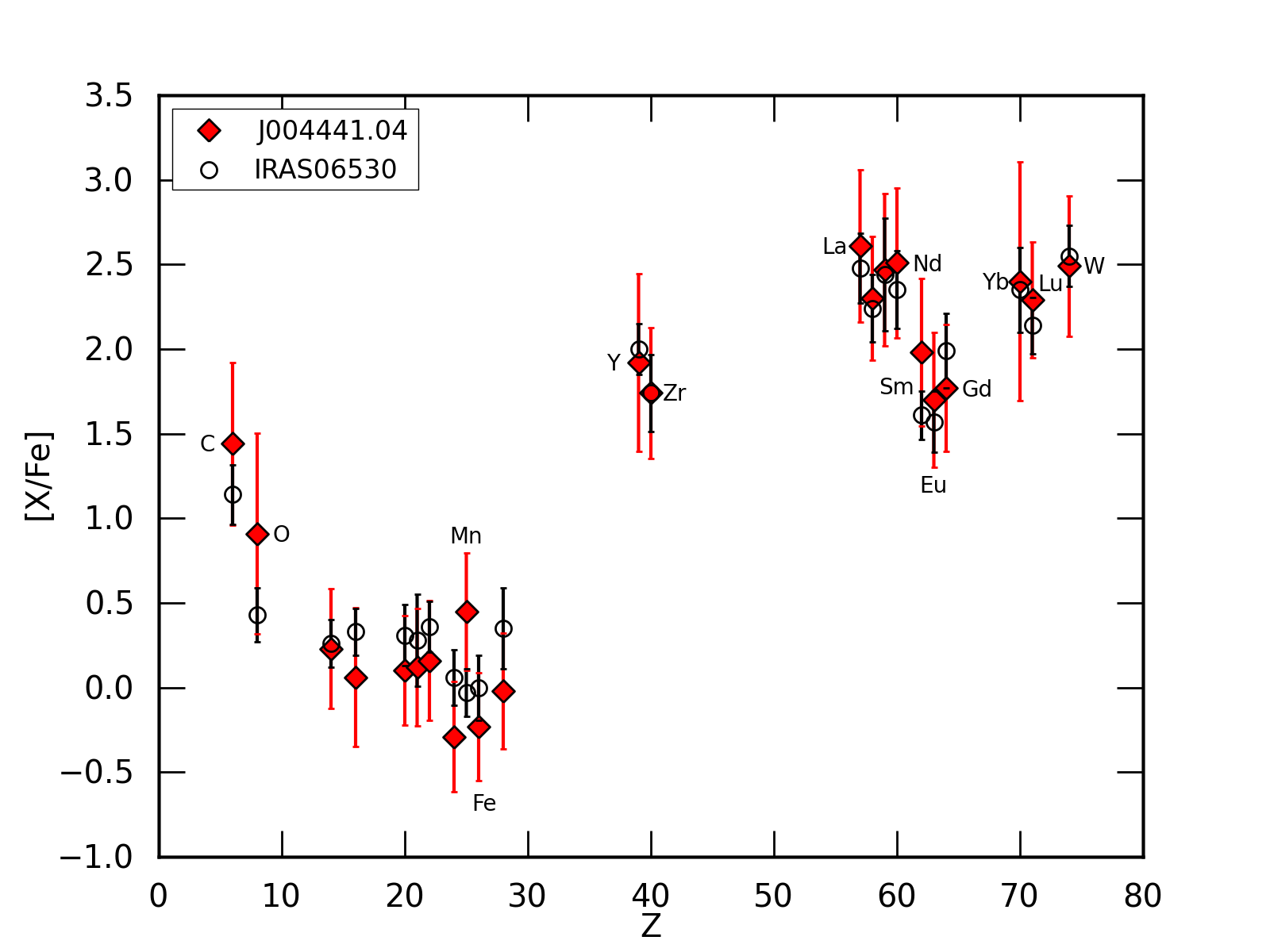}}
\caption{Comparison of $[\textrm{X/Fe}]$ results for J004441.04 and IRAS06530 \citep{reyniers04} scaled to $[\textrm{Zr/Fe}]$. Some elements are labelled for clarity. The red errorbars of J004441.04 ratios are the total uncertainty while for IRAS06530, the black errorbars represent line-to-line scatter. For more information, see text.}\label{fig:J00_vs_iras}
\end{figure}

To get a good comparative view of the s-process nucleosynthesis of
J004441.04 and IRAS06530, we scale the $[\textrm{Zr/Fe}]$ result of
J004441.04 to the $[\textrm{Zr/Fe}]$ result of IRAS06530 making both
$[\textrm{Zr/Fe}]$ values overlap. This difference between the
$[\textrm{Zr/Fe}]$ results is then added to the ratios of J004441.04.

The results of this scaling are shown in Fig. \ref{fig:J00_vs_iras} in
which the red diamonds are the $[\textrm{X/Fe}]$ results of J004441.04,
open circles represent the results of IRAS06530. The red errorbars of
J004441.04 ratios represent the total uncertainty in $[\textrm{X/Fe}]$
while for IRAS06530, the black errorbars represent only the
line-to-line scatter taken from \cite{reyniers04}. Some elements are
labelled for clarity. Despite a metallicity difference,
Fig. \ref{fig:J00_vs_iras} clearly shows that
both objects have very similar s-process abundance patterns.
With a C/O ratio of 2.8 and 1.9 respectively, also the C/O ratio of
IRAS06530 and J004441.04 are very similar.
The same abundance trends can be found
for the lighter elements where only Mn displays a strong difference.
Following the index definitions of \cite{reyniers04}, we find an
intrinsic index [hs/ls] = +0.5 and [s/Fe] = 2.4. In these indices, hs
stands for high-mass s-process elements around the Ba peak and ls for
elements around the Sr peak.  We quantified the Ba abundance 
using the observed Nd abundance. This  places J004441.04 in the upper right corner 
of the upper panel in Fig. 7 of \cite{reyniers04} confirming the correlation between 
the total enrichment of s-process elements and the [hs/ls] index. The results of J004441.04 
corroborate the finding  that there is no clear correlation between metallicity and the [hs/ls] 
index in the metallicity range sampled by post-AGB stars \citep{reyniers04}.

\section{Spectral energy distribution}\label{sect:SED}
The spectral energy distribution (SED) of J004441.04 and its assumed
distance provide the opportunity of determining its luminosity. 
The photometric data of Table
\ref{table:phot} are used for the SED construction of J004441.04. 
For post-AGB objects in the SMC, we need to take into account three possible
sources of reddening.

The first one is reddening by interstellar dust in the Milky Way
Galaxy (MW) towards the SMC. \citet{schlegel98} derived the extinction
towards the SMC by galactic dust to be small with an average of E(B-V) = 0.037 mag.
\begin{figure}[tb!]
\resizebox{\hsize}{!}{\includegraphics{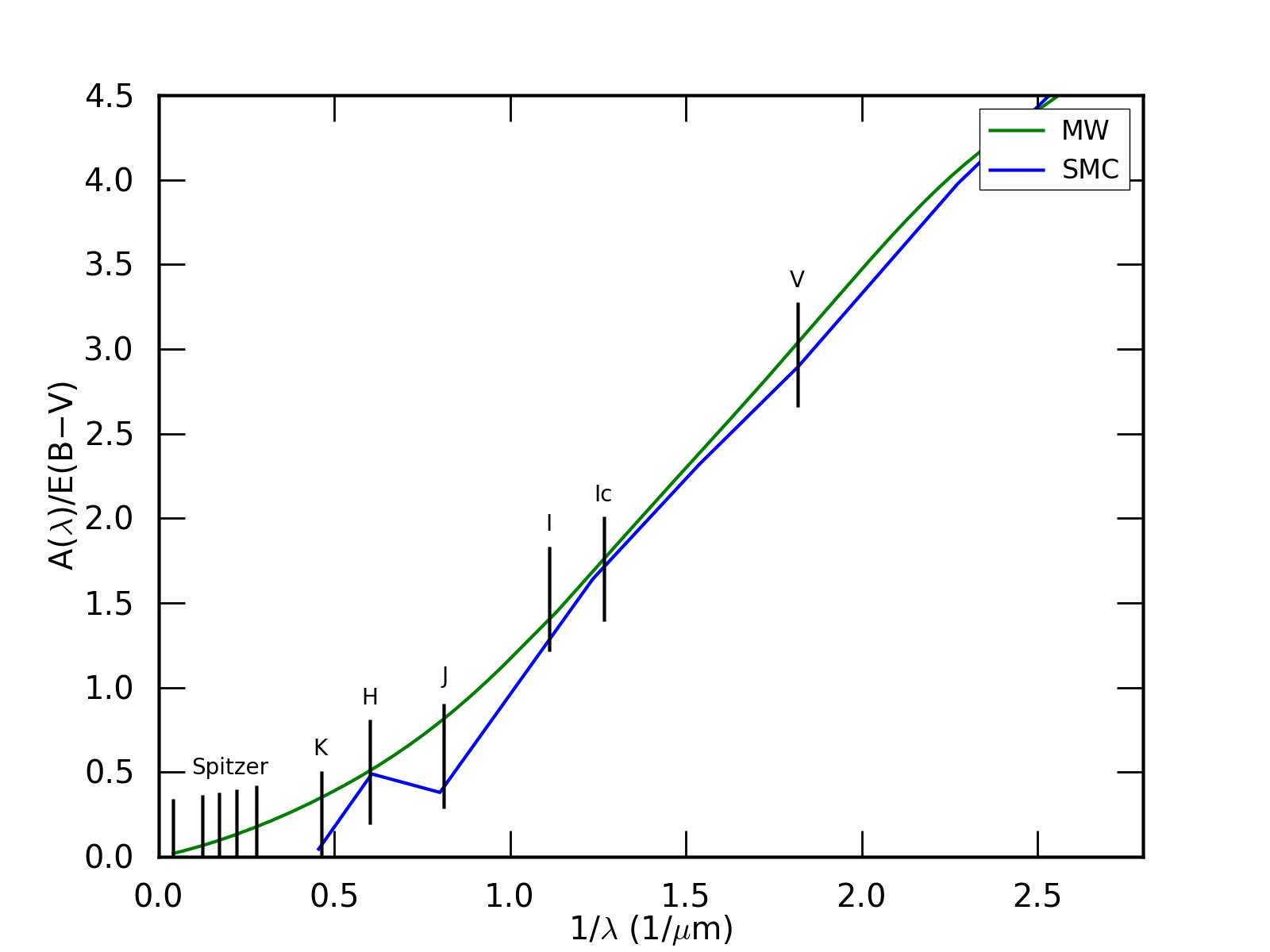}}
\caption{Extinction curves of the MW (green) and the SMC (blue). The position of the photometric bands used for the SED of J004441.04 are indicated. The MW extinction curve is taken from \citet{cardelli89}, the SMC extinction curve from \citet{gordon03}. The SMC curve only contains values at filter wavelengths explaining the strange behavior.}\label{fig:extinction_curve}
\end{figure}
\begin{figure}[t!]
\resizebox{\hsize}{!}{\includegraphics{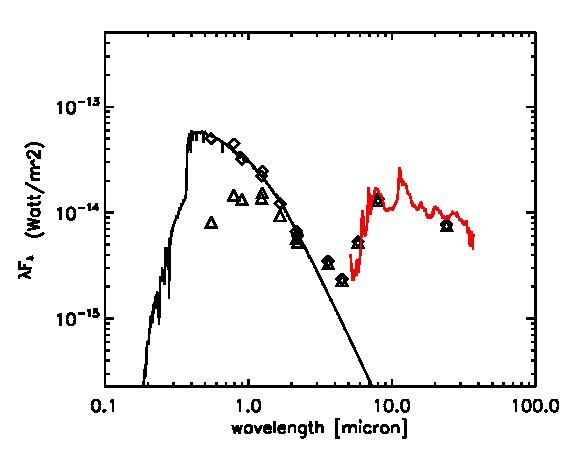}}
\caption{SED of J004441.04. Squares indicate the derived dereddened flux, triangles indicate original flux. The black line represents the Kurucz atmosphere model used to fit the photometry of J004441.04. The red spectrum is the IR spectrum which is used in \citet{volk11}.}\label{fig:sed}
\end{figure}

The second possible reddening source is
reddening in the SMC itself. Fig. \ref{fig:extinction_curve} shows the MW (green)
and SMC (blue) extinction curves which are taken from
\citet{cardelli89} and \citet{gordon03} respectively. The lines
indicate the extinction of the different photometric bands of Table
\ref{table:phot}. The SMC curve
only contains values at filter wavelengths explaining the strange
behavior. Since both J and K bands almost overlap, only one of them is
indicated in the figure. The extinction in the wavelength range used
for the SED of J004441.04 is approximately the same for both the MW
and the SMC. The largest difference between both reddening laws are
shown in the J and K bands. The extinction is small for these IR bands,
making that the different reddening laws hardly differ in the
wavelength regime of interest.  Because of
the strong resemblance between the MW and SMC curve in the used
wavelength range, we use the reddening law of the MW to determine the
reddening in the SMC.

The third extinction cause is reddening by the circumstellar dust
envelope of the post-AGB object itself. We applied a dereddening
assuming that the wavelength dependency of the circumstellar extinction
is similar to the ISM extinction law of the MW.

We determined the total reddening by applying a $\chi^2$ minimalisation on
the fit between the scaled model atmosphere and the dereddened
broadband fluxes. A Monte Carlo simulation is
used to determine the error on E(B-V) for 1000 arrays with a normal
distribution of the original flux. For J004441.04, this results in an
E(B-V) of 0.64 $\pm$ 0.02.

\begin{table}[tb!]
\caption{\label{table:sed} SED results of J004441.04 using a distance of 61 kpc.}
\begin{center}
\begin{tabular}{lc} \hline 
E(B-V) & 0.64 $\pm$ 0.02 \\
L ($\textrm{L}_{\sun}$) & 7600 $\pm$ 200 \\
$\textrm{L}_{IR}$/$\textrm{L}_{\star}$ &  0.34 $\pm$ 0.01 \\
\hline
\end{tabular}
\end{center}
\end{table}

The SED of J004441.04 is shown in Fig. \ref{fig:sed} together with the
IR spectrum of J004441.04 which was used in \citet{volk11} to identify
the object as a 21 $\mu$m source. Squares indicate the dereddened fluxes, triangles indicate
original (reddened) fluxes and the black line represents the scaled
atmospheric Kurucz model (see \ref{table:atmos}). 

J004441.04 has a clear double-peaked SED where the left peak represents the
stellar photosphere and the right peak is the IR excess created by
radiation of dust grains in the ejected AGB mass envelope
(Fig.~\ref{fig:sed}). 
By integrating the surface of the scaled Kurucz model, the luminosity
of J004441.04 can be obtained assuming the distance to the SMC is
approximately the distance to J004441.04. We use a distance of 61 kpc
\citep{hilditch05} to determine the luminosity which results in a
photospheric luminosity of 7600 $\pm$ 200 $\textrm{L}_{\odot}$ where
the luminosity error is again determined via Monte Carlo simulations.
The luminosity ratio of the IR excess and the
photosphere can give some information about the strength of the
circumstellar reddening.
The IR luminosity is simply determined by
integrating the surface of the IR excess. The resulting luminosity
ratio $\textrm{L}_{IR}$/$\textrm{L}_{\star}$ is 0.34 $\pm$ 0.01 so
about $34\%$ of the radiation emitted by the star is absorbed and
re-emitted by dust in the circumstellar envelope contributing to
reddening. We can assume the visible galactic extinction
$\textrm{A}_V$ can be approximated to $\textrm{A}_V \approx 3.1 \cdot
\textrm{E(B-V)}$ and via Fig. \ref{fig:extinction_curve} we can assume
the V band extinction of the MW and SMC are approximately the
same. The deduced E(B-V) of 0.64 $\pm$ 0.02 makes that roughly $20\%$ of the original
photosphere flux reaches the observer and the other $80\%$ of the
photosphere flux is absorbed or scattered outside the line-of-sight. 
Comparing this with the luminosity ratio
$\textrm{L}_{IR}$/$\textrm{L}_{\star}$ = 0.34 $\pm$ 0.01 indicates
that there is a significant contribution of the ISM extinction
in the SMC and/or that there is deviation from spherical symmetry in
the circumstellar shell.

\begin{figure}
\begin{center}
\resizebox{\hsize}{!}{\includegraphics{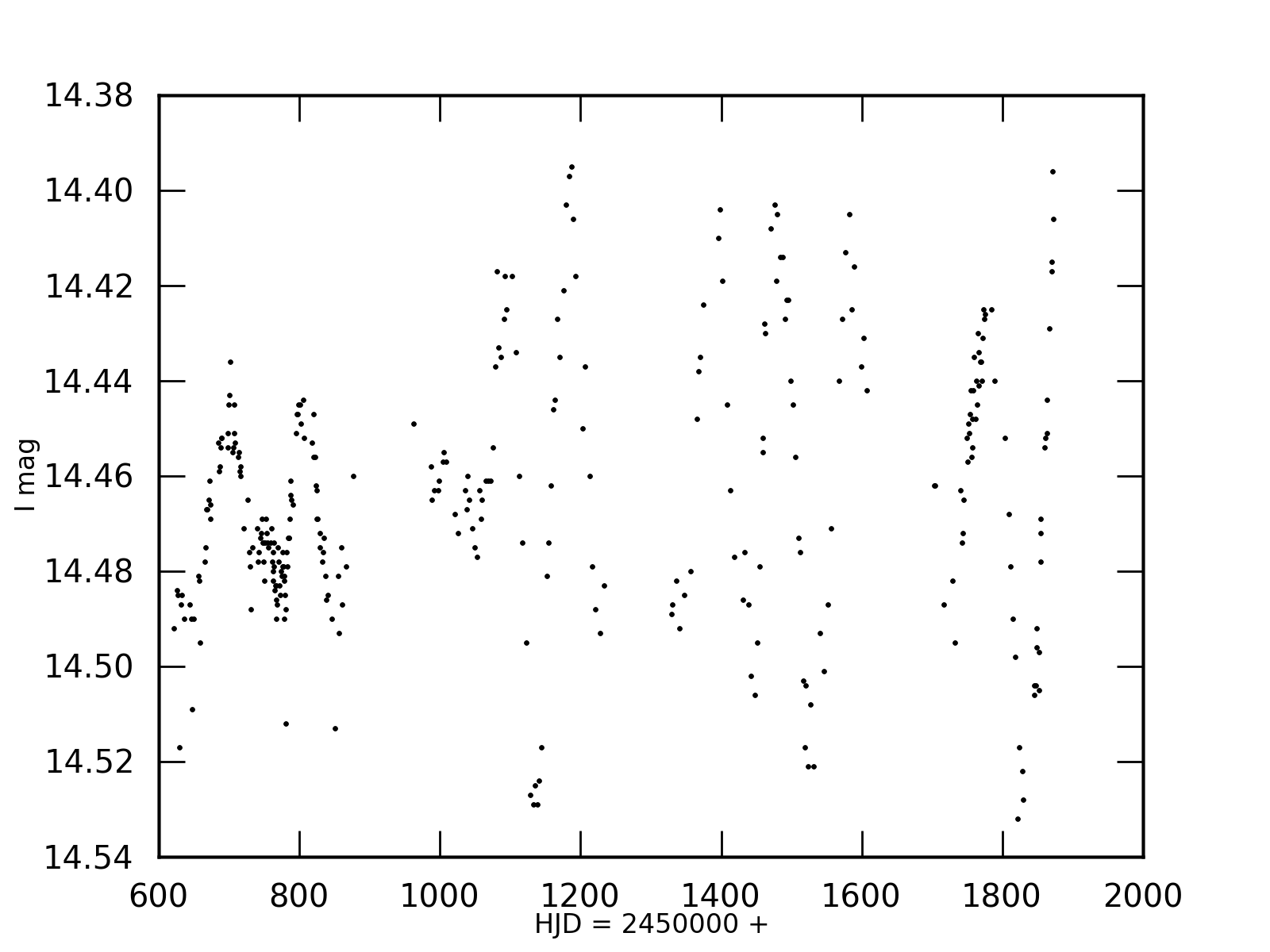}}
\caption{OGLE lightcurve of J004441.04 in the I band. For more information, see text.}\label{fig:lc}
\end{center}
\end{figure}

\begin{figure}
\resizebox{\hsize}{!}{\includegraphics{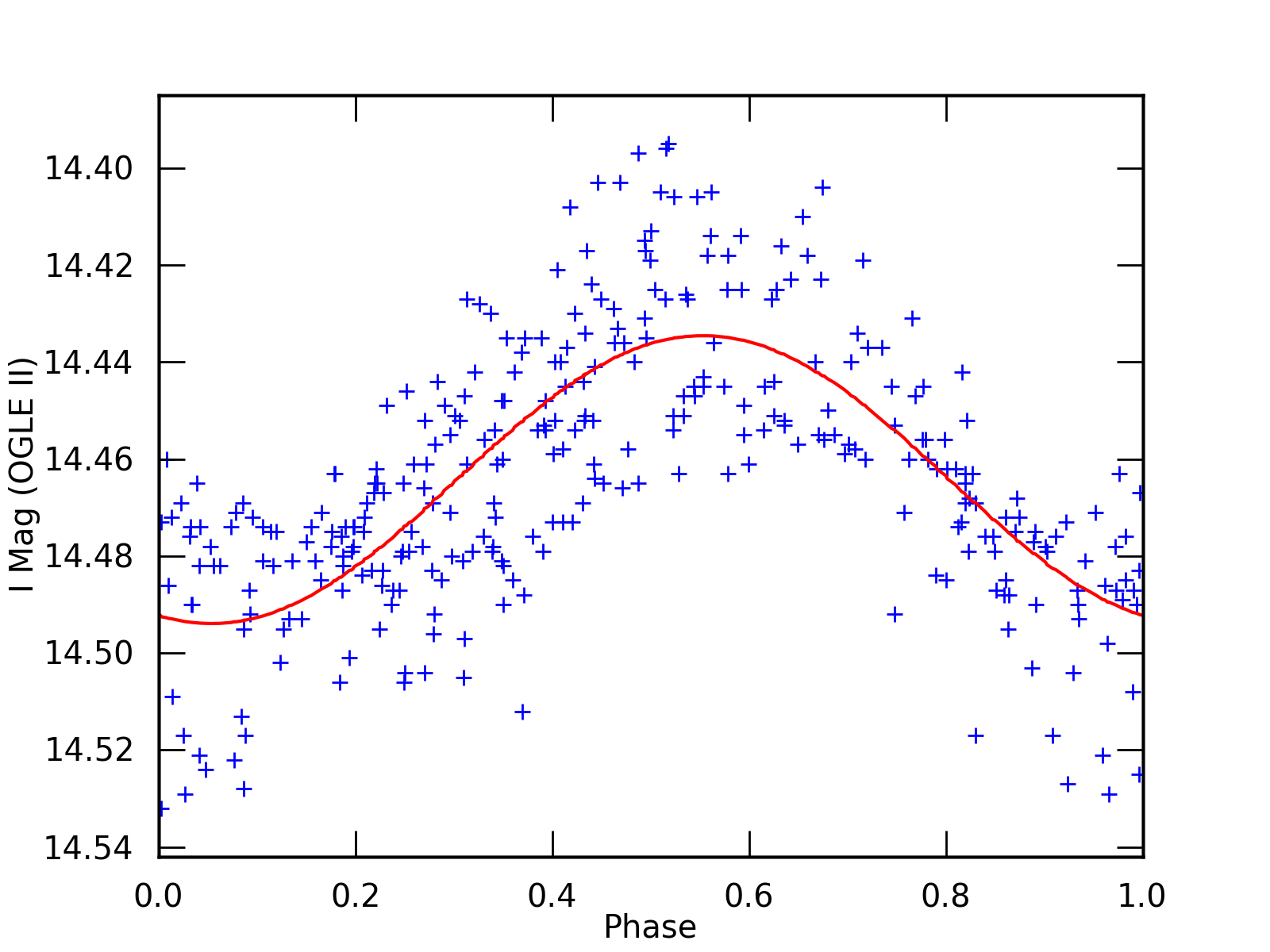}}
\caption{Phase diagram with the lightcurve of J004441.04 (blue crosses) folded over the dominant frequency of 97.6 $\pm$ 0.3 days (red line).}\label{fig:phase}
\end{figure}

Post-AGB stars pass through the Cepheid II instability strip and
J004441.04 is no exception. The variability in the I filter is
displayed in Fig. \ref{fig:lc}. A frequency
analysis of this light curve provides a dominating period of 97.6
$\pm$ 0.3 days which is plotted in a phase diagram in
Fig. \ref{fig:phase}. Blue crosses are the individual light
curve points of Fig. \ref{fig:lc} and it is clear from
Fig. \ref{fig:phase} that this is indeed the dominant period. 
Amplitude modulation is clearly apparent in Fig.~\ref{fig:lc}. As
the amplitude of the variability is rather small (maximum $\pm 0.15$
mag in I), the use of stable atmosphere models for the spectral analyses is
still appropriate.

We attempted to use the observed period of 97.6 days
to constrain the parameters of J004441.04 under the
assumption that this is a pulsation.  The linear 
pulsation code of \cite{wood76} and
its updates as described in \cite{kamath10} 
was used to examine the pulsation period of the fundamental
mode in stellar models with luminosities of 7600, 10000,
15000 and 18000 Lsun and $\textrm{T}_{\rm eff}$ values between 4700 and
6300 K.  The stellar mass was obtained from a linear
fit between L and M for the two post-AGB tracks of
Vassiliadis and Wood (1994) for an abundance of Z=0.001.
The composition of the models was assumed to be X=0.75
and Z=0.001.  

For our derived $\textrm{T}_{\rm eff}$ of 6250 K for J004441.04, the pulsation
period of all these models is too small, typically near 50 days. This
is about half the observed period.
A much cooler effective temperature is required in order to obtain
a model with the observed period of 97.6 days.  At the
luminosity L = 7600 L$_{\odot}$ (our estimated value), a fundamental
mode period of 97.6 days occurs at $\textrm{T}_{\rm eff} \sim 4830$ K while at
the highest luminosity investigated (18000 Lsun) a
fundamental mode period of 97.6 days occurs
at $\textrm{T}_{\rm eff} \sim 5010$ K.  In addition, models with $\textrm{T}_{\rm eff}$ near 6250 K
are pulsationally stable in modes with periods greater than 20 days while 
those with $\textrm{T}_{\rm eff}$ near 4900 K are unstable. 
The pulsation periods are relatively insensitive
to composition (X=0.10 was tried) and the adopted mass.

The $\textrm{T}_{\rm eff}$ value of $\sim$ 4900K required to get the correct pulsation 
period is, however, incompatible with the spectrum of J004441.04.
There are two possible explanations for
this discrepancy.  Firstly, it is possible that the pulsation
models are wrong.  These stars are truely extreme:
in the inner one third of the star radiation provides 90\% of 
the pressure supporting the exterior envelope.  The stars
are difficult to analyse but all the relevant physics is
included so there should be no obvious problem.  The metal abundances in the envelope
are non-standard so that the solar-scaled abundances used
will lead to errors in the opacity.  This is more likely
to affect the stability of the models than the periods.
It is not obvious that the pulsation models have significant incomplete physics.
 
If the pulsation models give the correct periods, then
a second explanation for the discrepancy is required.
Here we raise the possibility that the observed periodicity
is not a pulsation but it could be due to other surface phenomena like
rapid rotation of a spotted star or an orbiting companion.  With our adopted
$\textrm{T}_{\rm eff}$ and L values, and with a stellar mass of 0.65 $\textrm{M}_{\odot}$,
the breakup period is $\sim$ 94 days, very close to the observed
period.  So if rotation is an explanation for the variability,
the star must be rotating at near-breakup velocity, which is again
incompatible with the detected broadening of the spectral lines, unless
we happen to look along the rotation axes. The angle between the
line-of-sight and the rotation axes should, in that case,  be lower than 10$^{\circ}$.
Also, models indicate that a strong rotation and rotational shear in the intershell,
may mix $^{14}$N into the $^{13}$C pocket hence impact on the
efficiency of the s-process nucleosynthesis \citep{herwig03,siess04}.
Considering the other possibility, an orbiting companion or dust cloud would have 
to be very close to the surface of J004441.04. The SED shows no
indication of a hot dust excess (Fig.~\ref{fig:sed}). 
All these possibilities do seem very unlikely.  We are therefore unable
to provide a satisfactory explanation for the observed
periodicity in J004441.04.

\section{Initial mass determination}\label{subsect:inimass}
Before comparing the results of J004441.04 with theoretical
nucleosynthetic models, its initial mass (mass at the zero-age
main-sequence) needs to be estimated. Both metallicity and initial
mass are fundamental parameters for the calculation of models.

We use the evolutionary post-AGB tracks of \citet{vassiliadis94} to
obtain an accurate estimate of the initial mass. The choice of these
tracks is based upon the metallicity: these authors calculated tracks
for a metallicity of Z = 0.001 corresponding to the found metallicity
of J004441.04. Unfortunately, the \citet{vassiliadis94} tracks start
at an effective temperature of $\textrm{T}_{\rm eff} = 10^{4}$ K and hence
do not include the 6250 K of J004441.04. However, the luminosity of
post-AGB stars remains approximately constant during their transit
through the HR diagram, allowing linear extrapolation between log
$\textrm{T}_{\rm eff}$ and log L/$\textrm{L}_{\sun}$ towards lower
temperatures. The need for extrapolation to lower temperatures may
indicate that J004441.04 is in the beginning stage of the post-AGB
phase.
\begin{figure}
\resizebox{\hsize}{!}{\includegraphics{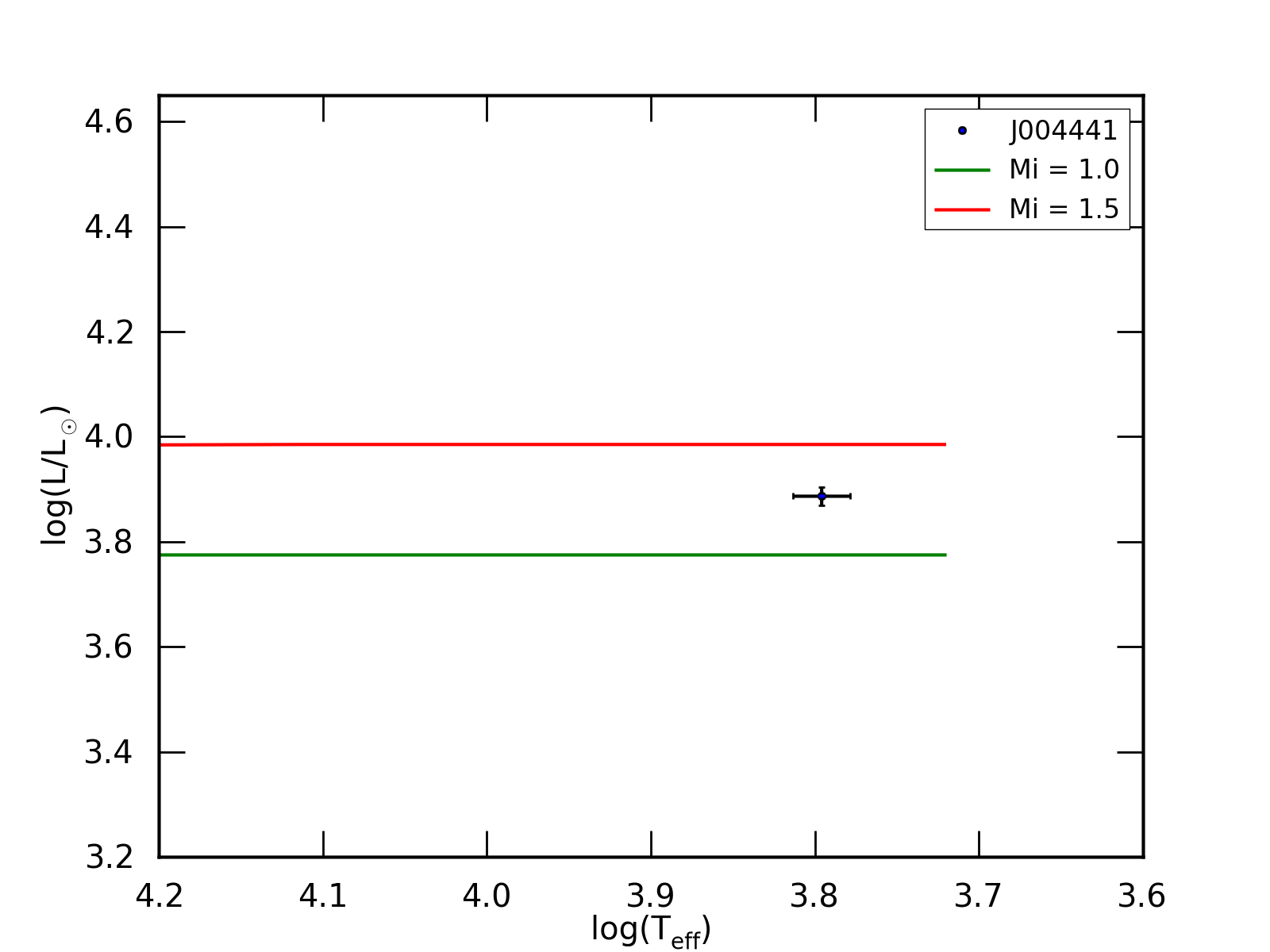}}
\caption{J004441.04 positioned in the HR diagram based upon the results in this paper. The red and green lines respectively represent the 1.5 $\textrm{M}_{\sun}$ and 1.0 $\textrm{M}_{\sun}$ evolutionary tracks of \citet{vassiliadis94} for Z = 0.001.}\label{fig:inimass}
\end{figure}

Fig. \ref{fig:inimass} shows the position of J004441.04 in the HR
diagram together with the 1.5 $\textrm{M}_{\sun}$ (red line) and 1.0
$\textrm{M}_{\sun}$ (green line) evolutionary tracks of
\citet{vassiliadis94} for Z = 0.001. Based upon this figure, we
estimate an initial mass of M $\simeq$ 1.3 $\textrm{M}_{\sun}$ for
J004441.04. The tracks for Z=0.001 predict
a mass for the current post-AGB star of 0.65 M$_{\odot}$.
With the derived T$_{eff}$ and L, the radius is 75 R$_{\odot}$,
and log g = 0.5 also gives M=0.65 M$_{\odot}$ which is consistent with
the tracks.

\section{AGB chemical models} \label{sect:models}

We compare the observed abundance results with predictions from two
independent stellar evolution codes, with which we calculated models of a
1.3M$_{\odot}$ star of [Fe/H] = -1.4.
\subsection{Mount-Stromlo Evolutionary predictions}

One of the AGB evolutionary models was calculated using the
same version of the stellar evolution code described in \citet[][and
references therein]{karakas10a}, which uses the \citet{vassiliadis93}
mass-loss rate on the AGB, and includes the addition of C and N-rich
low-temperature opacities tables from \citet{lederer09}.  Convective
overshoot is used to induce the third dredge-up (TDU) in the
1.3M$_{\odot}$, $Z= 0.0006$ model.  The TDU is the inward movement of
the convective envelope into regions processed by partial He-burning
during a thermal instability \citep[see][for a recent
review]{herwig05}.  Some form of overshoot is required because
low-mass models computed previously with the same code shows little or
no TDU for masses $\lesssim 2\,M_{\odot}$ at solar metallicity. We
calculate one stellar evolution sequence of a 1.3M$_{\odot}$ model
star with a scaled-solar initial abundance pattern with Z = 0.0006.

We include overshoot by extending the position of the base of the
convective envelope by 1.0 pressure-scale height.  We do not vary this
free parameter for this study, noting that the final predicted carbon
abundance of log $\epsilon$(C) = 8.90 is already 40\% higher than the observed
carbon abundance of log $\epsilon$(C) = 8.76.
The final predicted luminosity of the
stellar model at the tip of the AGB is 8400\,L$_{\odot}$, which is
achieved after 14 thermal pulses. The observed luminosity
of 7600\,L$_{\odot}$ is reached after 12 thermal pulses. 
It would be possible to match the observed luminosity at the
tip of the AGB by either increasing the mass-loss
rate or increasing the amount of overshoot. 

\begin{figure}
\begin{center}
\resizebox{\hsize}{!}{\includegraphics{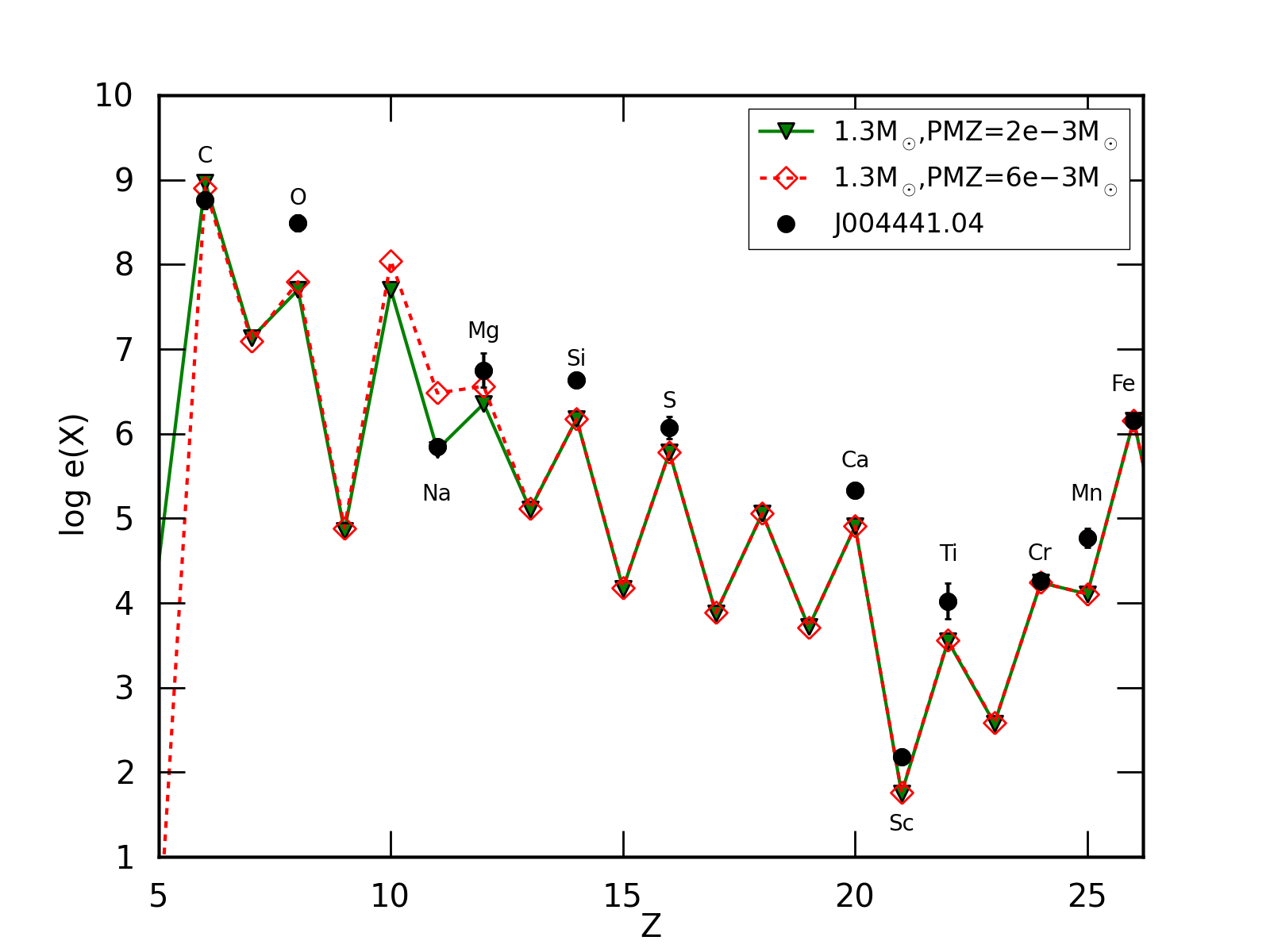}} 
\caption{The predicted abundance of elements lighter than
Fe, in $\log \epsilon$(X), as a function of atomic number, $Z$,
for a 1.3 M$_{\odot}$ model of $Z = 0.0006$ ([Fe/H] = $-1.4$). 
Abundances are sampled at the tip of the AGB, after the last
computed thermal pulse.
Included are the approximate locations (in proton number, $Z$) 
of some key elements. Predictions are shown for two
values of the parameter M$_{\rm mix}$ = 2 $\times$\, 10$^{-3}$ (points
connected by the solid green line) and 
6 $\times$ 10$^{-3}$\,M$_{\odot}$ (points connected by the red dashed line). 
This parameter determines the size of the $^{13}$C pocket
and $s$-process enrichment, see text for details. 
The derived abundance of the post-AGB star 
J004441.04 is shown with error bars.
\label{fig1}}
\end{center}
\end{figure}

\begin{figure}
\begin{center}
\resizebox{\hsize}{!}{\includegraphics{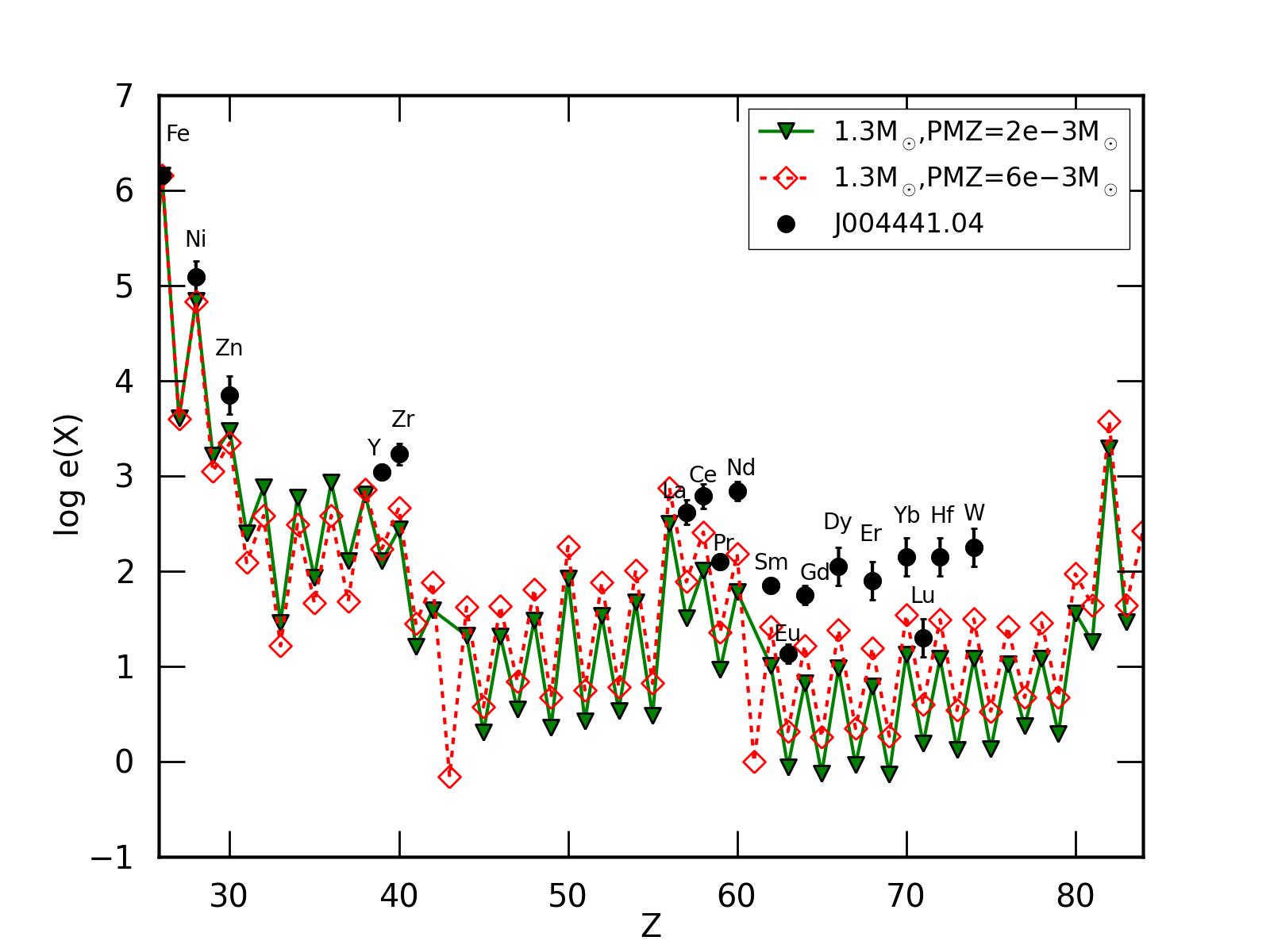}}
\caption{Same as Fig.~\ref{fig1} except for elements heavier
than Fe.
\label{fig2}}
\end{center}
\end{figure}

The $s$-process abundance predictions were calculated using 
the post-processing nucleosynthesis code and full network of 320 
species described in \citet{lugaro12}, with reaction
rates taken from the JINA REACLIB library \citep{cyburt10}.
For the initial composition we used the solar distribution of 
abundances from \citet{asplund09} scaled down to [Fe/H]=$-$1.4. 
Solar abundances of C, N, O, Ne, Mg, Si, S, Ar, and Fe are the 
pre-solar nebula values from Table~5 of \citet{asplund09}; F is 
the meteoritic value of log $\epsilon (F)_{\odot}$ = 4.42 
from Table~1 of the same paper (chosen because it has a lower uncertainty), 
and for many of the elements heavier than Fe we use the meteoritic values 
for the solar abundances (e.g., Sr, Eu, Pb). No $\alpha$-enhancement
was used at this stage.

In order to obtain an enrichment of $s$-process elements, we
artificially introduce some protons into the top of the
He-intershell. Note that this has become standard practice for the
simple reason that there is not enough $^{13}$C in the H-burning ashes
of AGB stars to make it an efficient neutron source.  Recent studies
investigating the formation of the $^{13}$C pocket by various
mechanisms have found that the proton abundance in the intershell
decreases monotonically \citep[e.g.,][]{cristallo09}.  We apply the
assumption that the proton abundance in the intershell decreases
monotonically from the envelope value of $\simeq$ 0.7 to a minimum
value of 10$^{-4}$ at a given point in mass located below the base of
the envelope. We do this in the same manner as described in
\citet{alvesbrito11} and \citet{kamath12}.

This {\em partially mixed zone} (or PMZ) 
is required in order to facilitate the formation of a 
$^{13}$C pocket which allows neutrons to be 
released by the$^{13}$C($\alpha$,n)O$^{16}$ reaction 
\citep{straniero95,gallino98,goriely00,herwig05}. The mass 
of the proton profile is a free parameter which we 
set as a constant mass. We adopt two choices for the 
extent of the partially mixed proton zone: 1) PMZ = 0.002M$_{\odot}$ (green
solid line in Figs.~\ref{fig1} and~\ref{fig2}),
and 2) PMZ = 0.006$\textrm{M}_{\odot}$ (red dashed line in Figs.~\ref{fig1}
and~\ref{fig2}).

In Figs. \ref{fig1} and~\ref{fig2} we show the abundances derived
for J004441.04 in this work along with nucleosynthesis predictions
from this model. The main point of this comparison is to show that a
standard, initial scaled-solar abundance pattern used in the
stellar model produce $s$-process overabundances that are
{\em not} compatible with the derived abundances for 
J004441.04. The main shortcomings are that: 
\begin{itemize}
\item  While
  the predicted C overabundance is only 40\% higher than the observed
  value, the predicted O overabundance is too low. This makes that the
  C/O prediction is $\sim$18, which is clearly in contrast to the
  detected value of 1.9$\pm$0.7. An initial alpha enhanced 
  enrichment of [O/Fe] = +0.4, still makes the final O abundance too
  small and the high [O/Fe] = 1.14 indicates that
  the star has dredged-up a considerable amount of oxygen, above the
  level found in the stellar model of [O/Fe] = 0.35~dex. 
\item The total overabundances of the s-process nuclei are clearly too low;
\item While the predicted s-process {\sl distribution} is very
  similar to the observed one, the predicted  Pb abundance is
  significantly higher than the observed upper limit (see Sect. 
  \ref{subsect:Pbdis} and Fig. \ref{fig:Pbcomp}).
\end{itemize}

Note also that the final predicted $^{12}$C/$^{13}$C ratio is
extremely high at $\approx 1800$. An observational constraint on the
isotopic ratios is not possible with the wavelength coverage we have 
at present, but this should be certainly a priority in our next
observational season.

It is plausible that at the low metallicity of [Fe/H] = $-1.34$, 
J004441.04 evolved from an initially $\alpha$-enhanced composition.
This would explain the discrepancy between the predicted and observed 
Si, S and Ca abundances.

\subsection{STAREVOL}

To test the uncertainties associated with stellar modeling, an
additional 1.3$\textrm{M}_{\odot}$, 
{[}Fe/H{]}=-1.4 model was computed with the STAREVOL code
\citep[e.g.][and references therein]{siess07}. The initial
composition is slightly different, scaled solar according to
\cite{cunha06} which is based upon the \cite{asplund05} composition with
neon enhanced by nearly a factor of 2. This change in the neon abundance
was required to reproduce a correct seismic solar structure. As before, we
do not consider $\alpha$- enhancement and, with our reference composition,
the star has a metallicity Z=0.0044.  We use the Reimers mass loss rate
with $\eta=0.5$ up to the beginning of the AGB phase and then switch to the
\cite{vassiliadis93} prescription. The effect of CO opacities is
modeled using the analytical fits described in \cite{marigo02}. 
Overshooting is also included at the base of the convective
envelope but no overshoot from the flash-driven convection zone into
the CO core is taken into account. We use the diffusive approach of \cite{herwig99} with
$f_{over}=0.02$. Similar to the Stromlo model, this AGB model
experiences 15 thermal pulses and the TDU
starts at pulse number 6.  Considering the small overshoot parameter, the
amount of dredged-up material is rather limited and, except for pulse
number 9, $\lambda\lesssim0.06$ (where $\lambda$ is the ratio of the dredged-up mass
to the intershell mass increase between 2 consecutive pulses). At the tip
of the AGB, the stellar luminosity is 9400\,$L_{\odot}$. 
As previously
mentioned, this value can be lowered if a higher mass loss rate is
used. 

The STAREVOL network includes 55 species and allows an accurate
treatment of the nucleosynthesis up to $^{37}$Cl (for details, see \cite{siess08}). In this code, mixing and nuclear burning are
solved simultaneously, once the structure has converged. For the s-process
nucleosynthesis, we use the post-processing code as described in
\cite{goriely01a} which includes about 547 nuclei up to Po
with all relevant nuclear (n-, p-, $\alpha$-captures), weak (electron
captures, - decays) and electromagnetic (photodisintegration)
interactions. Nuclear reaction rates are taken from the updated Nuclear
Astrophysics Library of the Brussels University (BRUSLIB, available at
\texttt{http://www.astro.ulb.ac.be/bruslib}).  Following
\cite{goriely00}, a
PMZ of about half the mass extension of the pulse (i.e some
$6-10\times10^{-3}\textrm{M}_{\odot}$) is introduced below the convective envelope
at the time of the TDU.  In this PMZ, the proton profile decreases
exponentially from the envelope mass fraction down to $10^{-6}$ allowing
for the formation of a substantial $^{13}$C pocket.

\begin{figure}
\includegraphics[width=1\columnwidth]{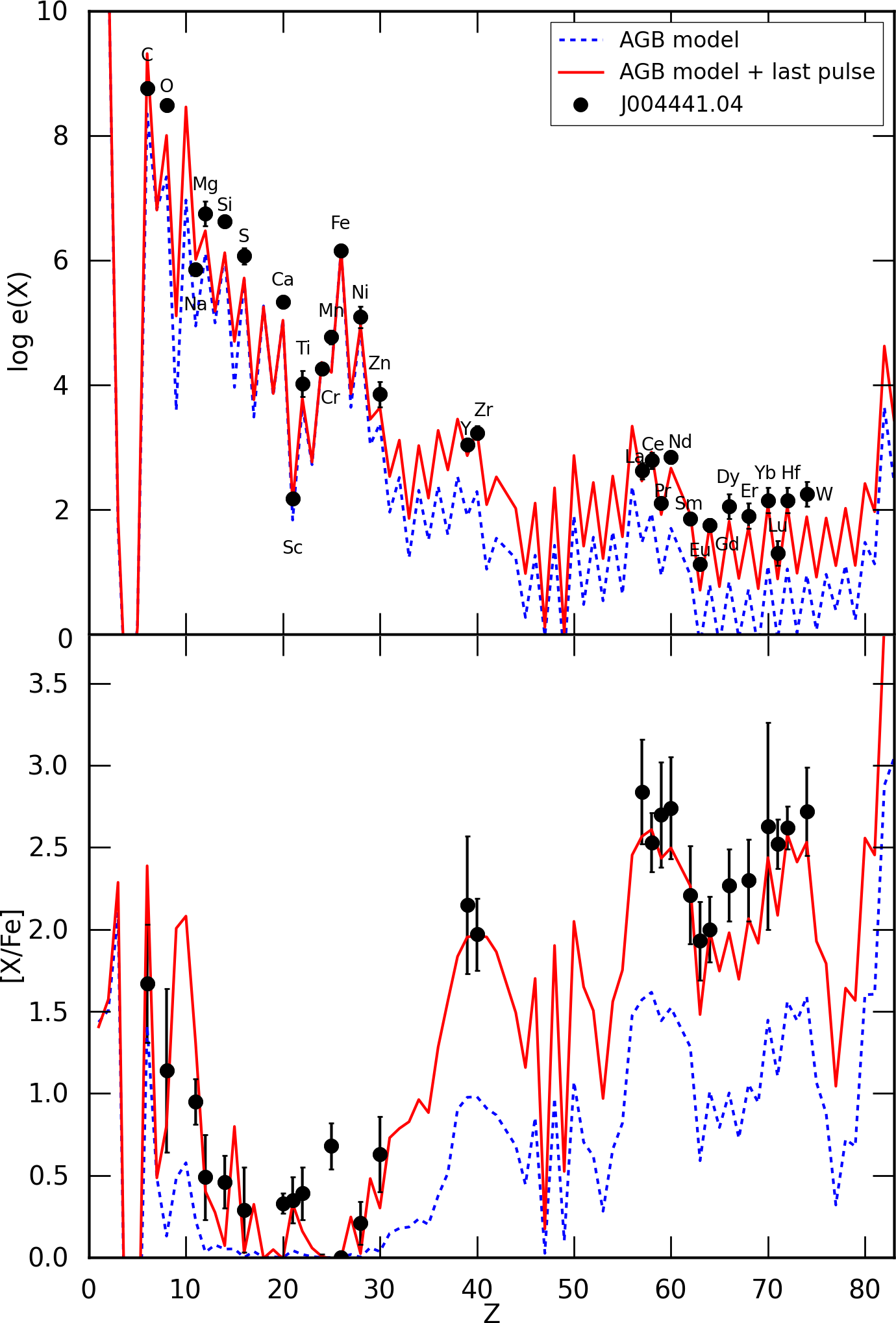}
\caption{Predicted elemental distribution in log $\epsilon$ (top) and {[}X/Fe{]}
(bottom panel) as a function of the atomic number Z for a 1.3$M_{\odot},${[}Fe/H{]}
= -1.4 model computed with STAREVOL. In each panel, the blue-dashed line corresponds
to the post-processed AGB model including 15 thermal pulses and the
red line shows the results expected from the occurrence of an additional final thermal
pulse followed by a deep TDU. \label{fig:starevol}}
\end{figure}

The results of the post-processing calculations are shown by the blue line
in Fig.\ref{fig:starevol}. The model has the same basic shortcoming, namely an oxygen
enrichment ([O/Fe] =0.13) too weak by a factor of $\sim$10 but a 
carbon enrichment ([C/Fe]=1.42) compatible with the observed value.
As illustrated in Fig.\ref{fig:starevol}, the surface s-process enrichment is also too low with
respect to the observations. The s-process abundance pattern itself is
in good agreement except again for Pb (see Sect. \ref{subsect:Pbdis}).

At the end of the AGB evolutionary phase, we are left with a relatively small 
convective envelope of less than 0.05 $\textrm{M}_{\odot}$. If we assume that our star model experiences 
a very last thermal pulse with a deep TDU, a substantial surface pollution follows and a 
good agreement with the observations can be achieved (red-solid line in Fig.18). This 
last-pulse modeling corresponds to a post-processing calculation identical to those 
characterizing the previous TDU episodes but for which the TDU is just assumed to extend 
to about $7 \times 10^{-3}$ $\textrm{M}_{\odot}$ deep into the C-rich region, i.e about half of the pulse mass 
is now diluted into the envelope. This improved fit to the observation is due
to the small remaining envelope mass ($M_{env}<0.05M_{\odot})$ which
limits the dilution of the dredged-up material. The observed
distribution of s-process elements can be nicely reproduced except for
Pb which, in both simulations, is overproduced (see Sect. \ref{subsect:Pbdis}), as classically done in
all s-process simulations at low-metallicity \citep{goriely00}.
Concerning the light elements, this last dredge-up brings
significantly more C, O, F, Ne and Na to the surface. The C is now in
disagreement with the observations but on the other hand, we can
reproduce the oxygen enrichment, yielding a similar C/O$\sim20$ as
before. We also note that the underabundance of the $\alpha$-elements
Si, S and Mn remains.  It is interesting to see how the use of
different abundances units, namely $\log(\epsilon)$ and {[}X/Fe{]} can
be misleading.
Finally, note that within the s-process model adopted here, the partial mixing of protons 
into the C-rich region also leads to a significant production of F, Ne and Na (see e.g \cite{goriely00})
which is also the case for the Mount-Stromlo models as shown in Fig. \ref{fig:modelcompare}. 
The estimated abundance of Na by the models is seen in Fig. \ref{fig:modelcompare} to be in rather good 
agreement with observation. It would be extremely valuable to confirm observationally the predicted 
F overproduction of about 2 dex. Unfortunately, at these photospheric
temperatures, F has no suitable lines in the sampled spectrum and its abundance can not be quantified. 

\begin{figure}
\includegraphics[width=1\columnwidth]{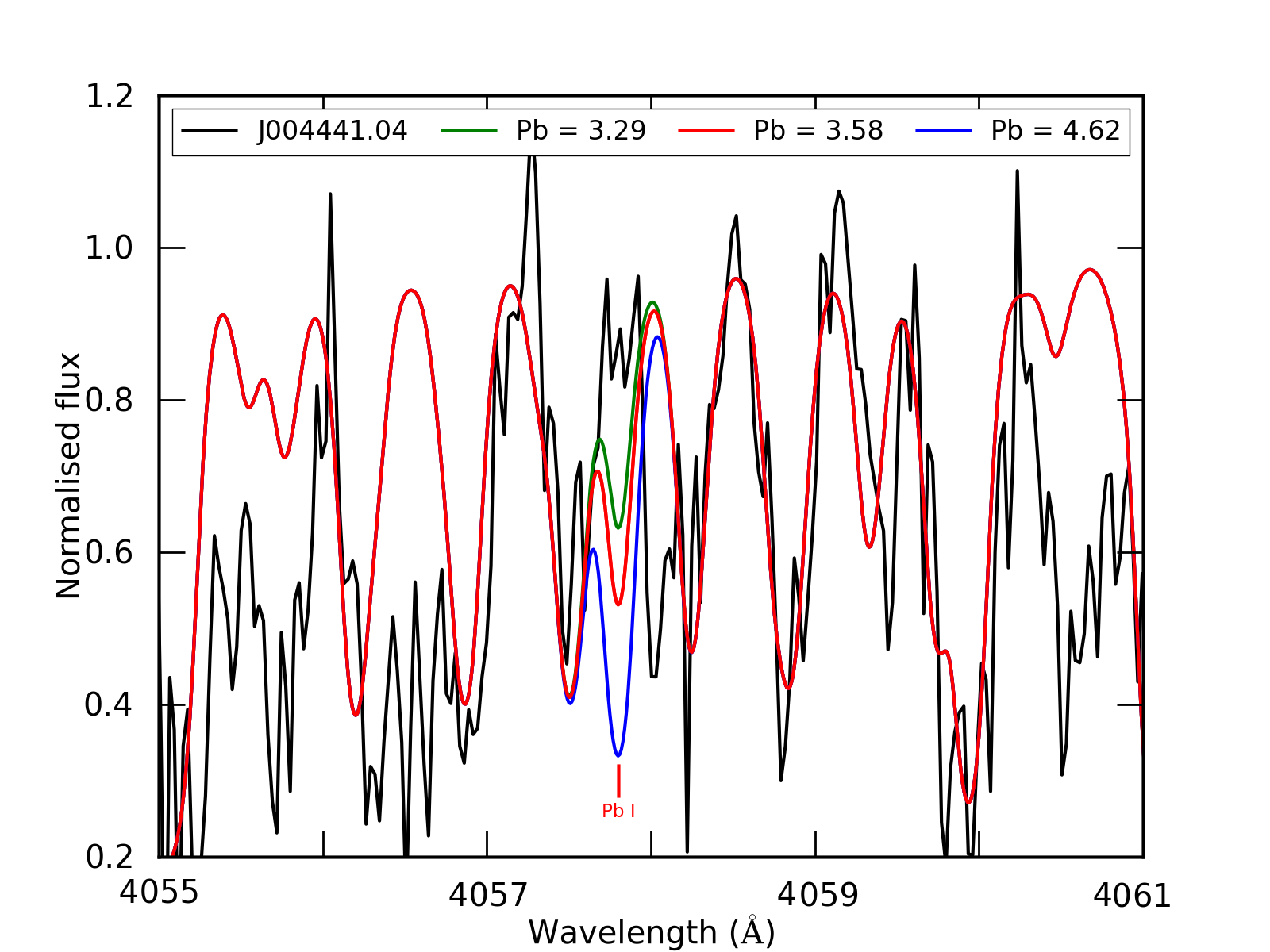}
\caption{ Pb I line region in the spectrum of J004441.04 compared with
 model predictions. In green, the Mount Stromlo model prediction (Section
 6.1) and in red and blue the predictions obtained by the STAREVOL code
 (Section 6.2). \label{fig:Pbcomp}}
\end{figure}

\subsection{Pb discrepancy}\label{subsect:Pbdis}

Figs. \ref{fig2}, \ref{fig:starevol} and \ref{fig:modelcompare} clearly show that current
stellar evolution codes predict strong Pb overabundances for
J004441.04. The strongest predicted Pb line in our sampled spectral
range is 4057.807 \AA{}. 
Unfortunately, this spectral region has a low S/N preventing an accurate 
Pb abundance determination. We focused on this spectral region and
looked carefully on the three exposures. We obtained a weighted
average of the three UVES spectra and 
then confronted this final spectrum with spectral synthesis models. We 
used the Pb abundances predicted by the chemical
models in Fig. \ref{fig:Pbcomp}. The black spectrum is the normalised weighted mean spectrum of J004441.04,
the green spectrum displays the predicted Pb abundance of the Mount-Stromlo model with PMZ = 
$0.006M_{\odot}$, the red and blue spectrum represent the STAREVOL predictions with the blue
spectrum the model with a last thermal pulse. Although the stellar
spectrum is of poor quality, it is clear from Fig. \ref{fig:Pbcomp} that the predicted strong
line is not present in the stellar spectrum. From Fig. \ref{fig:Pbcomp} we estimate an upper Pb 
abundance of 3.00 for J004441.04. Unfortunately we do not have a
spectrum of IRAS06530 in this spectral domain.
In the late thermal pulse scenario (Fig.~\ref{fig:modelcompare}), the
predicted Pb abundance becomes very large, which is
incompatible with the detected spectrum.  To reduce the production of Pb, 
the neutron in the PMZ production must be reduced. Partial pollution 
of the $^{13}$C pocket by $^{14}$N induced by rotational mixing 
\citep[e.g.][]{herwig03,siess04} may be needed to solve this problem 
but this will need a detailed investigation which is outside the scope 
of this paper.

\section{Discussion and Conclusion}\label{sect:conclusion}

We presented a detailed analysis of J004441.04, the only known 21
$\mu$m object of the SMC. This metal poor star ([Fe/H] = $-$1.34 $\pm$
0.32), turns out to be among the most s-process enriched stars known
to date while displaying only a moderate C/O ratio of 1.9 $\pm$ 0.7.
The atmospheric parameters (T$_{eff}$ = 6250 $\pm$ 250, log g = 0.5
$\pm$ 0.5) combined with the luminosity of 7600 $\pm$ 200
$\textrm{L}_{\odot}$ as well as the pulsation period of 97.6 $\pm$ 0.3
days, show that we witness the post-AGB phase of a star of low
initial mass. This initial mass is estimated to be $\simeq$\,1.3
M$_{\odot}$. Our findings on an extra-Galactic source corroborate the
conclusion that 21 $\mu$m stars are post-Carbon stars. Although of
lower metallicity, the object displays very similar overabundances as
the most s-process rich post-AGB star known in our Galaxy (IRAS06530-0213).

\begin{figure}
\includegraphics[width=1\columnwidth]{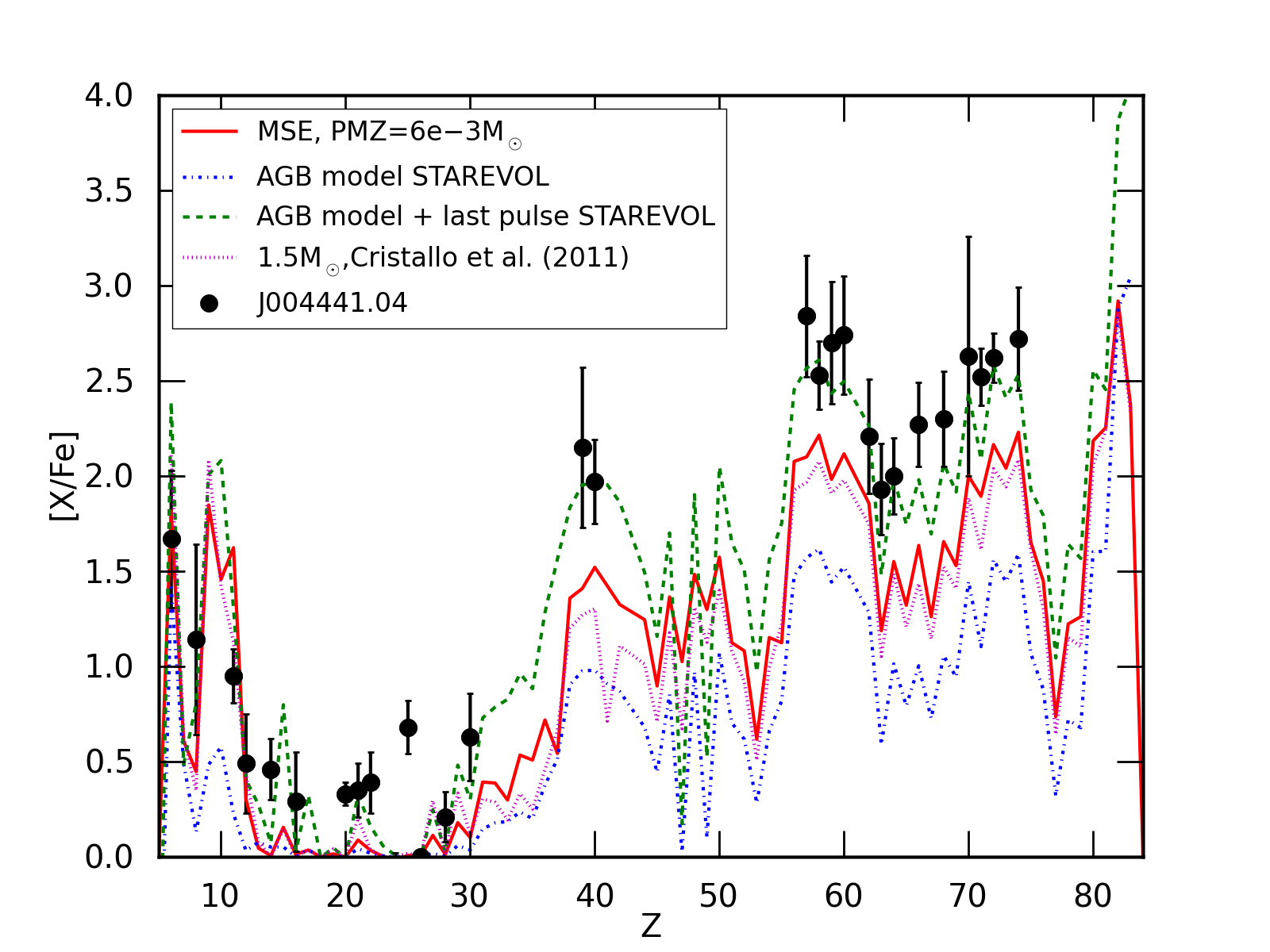}
\caption{ The abundance pattern of J004441.04 in comparison with the
 model predictions. In red, the Mount Stromlo model (Section
 6.1), in blue the predictions obtained by the STAREVOL code (Section 6.2), the green dashed 
 curve is computing assuming a late thermal pulse with deep dredge-up occurs when 
 the dilution is small. The magenta line is the 1.5 M$_{\odot}$, Z = 0.001 model of \cite{cristallo11}. \label{fig:modelcompare}}
\end{figure}

Our AGB model predictions are based on model calculations which
include a forced overshoot of the convective boundary as well as an
artificial inclusion of a proton profile into the intershell. We used
two independent codes for these predictions. 
For comparsion, we also included in Fig.~\ref{fig:modelcompare} the nearest
model in the coarser grid of models by \cite{cristallo11} (1.5 M$_{\odot}$, 
Z = 0.001) which yields similar discrepancy like our two models. 

While the predicted C overabundance is
compatible with the observations, the predicted O abundance is
significantly lower resulting in a predicted C/O
ratio of $\sim$20 which is clearly too high. 
More oxygen could be dredged-up in the models, if one includes additional
overshooting below the thermal pulse, as shown by \cite{herwig00}.

While the models fit the s-process
distribution well, as depicted in Fig.~\ref{fig:modelcompare}, the absolute abundances 
are not well matched. In our late thermal pulse scenario, we assume
that a thermal pulse occurred with dredge-up and a limited dillution by the
remaining reduced envelope mass. In this model, the very high overabundances are
matched better. A noticeable exception is that the 
predicted Pb abundance is much larger than the detected upper limit (Fig. \ref{fig:Pbcomp}). 
A higher S/N spectrum in the region of Pb lines is needed
to better quantify the Pb abundance which would yield an ever stronger
test on the nucleosynthesis but our spectrum is incompatible with such
high Pb abundance.
With the data at hand, no isotopic ratio could be determined yet but
the predicted $^{12}$C/$^{13}$C of $\sim$ 1800 is extremely high.
The nearest model in the coarser grid of models by \cite{cristallo11}
yields similar discrepancy with even a higher predicted C/O ratio and  $^{12}$C/$^{13}$C values.

Our calculations demonstrate
that there is only a weak dependency of the theoretical predictions
on the adopted stellar evolution code !

It is interesting to compare the abundance pattern of J004441.04 with
the patterns observed in peculiar metal poor stars like the CEMP-$s/r$
stars, that is, a carbon-enhanced metal-poor star with enrichments in
both $s$ and $r$-process elements \citep[see discussions
in][]{beers05,jonsell06,sneden08,lugaro12}.  According to the
definition given in \citet{jonsell06}, CEMP-$s/r$ stars have
[Eu/Fe]$>$1 and [Ba/Eu]$>$0. The post-AGB star J004441.04 satisfies
all of these criteria. It is carbon rich with [C/Fe] = 1.67 and
[Ba/Eu] $> 0$, using the [Ba/Fe] = 2.67 estimated earlier. The
estimated Ba (or observed La, which is a good proxy for elements at
the second $s$-process peak) and Eu abundances put this star well into
the region of the [Eu/Fe]--[Ba/Fe] plane occupied by the CEMP-$s/r$
stars \citep[see Fig.~5 from][]{lugaro12}. Abundance wise, J004441.04
could be seen as a more metal rich analogue of the CEMP-$s/r$ stars.
However, unlike many of the stars within the CEMP-$s/r$ class, the
distribution of heavy elements in J004441.04 is similar to that of a
pure s-process, except for the anomalously low Pb abundance. What is
unusual about J004441.04 is how the star obtained such large
enrichments of heavy elements, while keeping the C/O ratio low.

It is fair to say that many of the chemical, physical and pulsational
properties of the interesting SMC post-AGB star J004441.04 are, as yet, unaccounted 
for by our model predictions.
This study is the first of a series in which we aim at a systematic
study of post-AGB stars in the Magellanic clouds. These objects with
constrained distances will provide unprecedented and systematic tests of the AGB
model predictions, complementary to the limited study of AGB stars
\citep[e.g.][]{delaverny06,abia11} for which much lower overabundances
are detected. The ultimate aim of this programme is to make progress
in the understanding of the complex interplay between mixing,
nucleosynthesis and mass-loss which characterise the final evolution
of solar-like stars.

\begin{acknowledgements}
  HVW and KDS acknowledge financial support from the Research Council of
  K.U.Leuven under grant GOA/2008/04 and from the Scientific Fund of
  Flanders (FWO) under the grants G.0703.08 and G.0470.07. 
AIK is grateful for the support of the NCI National Facility at the
ANU. AIK is an ARC Future Fellow and is supported under grant
FT10100475. LS and SG are FNRS
research associates. The authors thank Chris Sneden for his comments on the manuscript. 
\end{acknowledgements}

\bibliographystyle{aa}
\bibliography{allreferences}

\onecolumn
\begin{longtable}{cccrrr}
\caption{\label{tab:all_lines} Linelist used for the abundance determination of J004441.04. The right column shows the calculated equivalent width. EWs indicated with SS are calculated via synthetic modelling.} \\
\hline\hline
ion & Z & wavelength (\AA{}) & EP(eV) & log gf & EW (m\AA{}) \\ 
\hline
\endfirsthead
\caption{continued.}\\
\hline\hline
ion & Z & wavelength (\AA{}) & EP (eV) & log gf & EW (m\AA{}) \\ 
\hline
\endhead
\hline
C I  & 6  & 6655.517 & 8.54 & -1.94 & 56.0 \\
  &  & 6012.225 & 8.64 & -2.00 & 48.8 \\
  &  & 6014.834 & 8.64 & -1.59 & 71.0 \\
  &  & 6016.444 & 8.64 & -1.83 & 74.6 \\
  &  & 5551.578 & 8.64 & -1.90 & 53.2 \\
  &  & 6002.987 & 8.65 & -2.17 & 39.3 \\
\hline
O I  & 8  & 6155.971 & 10.74 & -0.67 & 18.7 \\
  &  & 6454.444 & 10.74 & -1.08 & 9.7 \\
  &  & 6453.640 & 10.74 & -1.30 & 7.3 \\
\hline
Na I  & 11  & 6154.230 & 2.10 & -1.56 & 11.4 \\
  &  & 5688.210 & 2.10 & -0.42 & 74.7 \\
\hline
Mg I  & 12  & 5711.100 & 4.34 & -1.75 & 38.0 (SS)\\
\hline
Si I  & 14  & 6237.330 & 5.61 & -0.53 & 35.6 \\
  &  & 5948.550 & 5.08 & -1.22 & 22.5 \\
\hline
S I  & 16  & 6757.160 & 7.87 & -0.20 & 20.7 \\
  &  & 6748.790 & 7.87 & -0.35 & 10.5 \\
\hline
Ca I  & 20  & 6102.727 & 1.88 & -0.80 & 59.7 \\
  &  & 6449.820 & 2.52 & -0.62 & 33.3 \\
  &  & 5590.130 & 2.52 & -0.74 & 26.6 \\
  &  & 6439.075 & 2.53 & 0.39 & 116.4 \\
\hline
Sc II  & 21  & 6604.600 & 1.36 & -1.23 & 98.6 \\
  &  & 5552.240 & 1.45 & -2.08 & 15.3 \\
  &  & 5667.153 & 1.50 & -1.21 & 68.3 \\
\hline
Ti II  & 22  & 4798.521 & 1.08 & -2.67 & 84.5 \\
  &  & 6606.980 & 2.06 & -2.85 & 29.0 \\
  &  & 5010.212 & 3.10 & -1.30 & 76.7 \\
\hline
Cr II  & 24  & 5510.730 & 3.83 & -2.48 & 18.2 \\
  &  & 5310.700 & 4.07 & -2.28 & 15.8 \\
  &  & 5502.090 & 4.17 & -1.99 & 22.3 \\
\hline
Mn I  & 25  & 6021.800 & 3.07 & 0.03 & 39.6 \\
  &  & 6016.650 & 3.07 & -0.22 & 33.1 \\
\hline
Fe I  & 26  & 5198.711 & 2.22 & -2.14 & 33.3 \\
  &  & 5049.820 & 2.28 & -1.35 & 80.7 \\
  &  & 6421.351 & 2.28 & -2.01 & 33.4 \\
  &  & 6252.555 & 2.40 & -1.72 & 41.7 \\
  &  & 6191.558 & 2.43 & -1.42 & 60.1 \\
  &  & 6136.596 & 2.45 & -1.40 & 65.5 \\
  &  & 6230.723 & 2.56 & -1.28 & 59.2 \\
  &  & 6137.692 & 2.59 & -1.40 & 47.5 \\
  &  & 5281.790 & 3.04 & -0.83 & 76.9 \\
  &  & 5324.179 & 3.24 & -0.10 & 108.5 \\
  &  & 5615.644 & 3.33 & 0.05 & 118.6 \\
  &  & 6400.001 & 3.60 & -0.29 & 66.2 \\
  &  & 6411.649 & 3.65 & -0.66 & 37.7 \\
  &  & 5445.042 & 4.39 & 0.04 & 40.0 \\
  &  & 6024.058 & 4.55 & -0.06 & 29.5 \\
  &  & 6419.950 & 4.73 & -0.09 & 15.3 \\
  &  & 5572.842 & 3.40 & -0.28 & 91.9 \\
\hline
Fe II  & 26  & 5991.368 & 3.15 & -3.56 & 54.7 \\
  &  & 6084.099 & 3.20 & -3.80 & 48.5 \\
  &  & 6147.741 & 3.89 & -2.73 & 66.1 \\
  &  & 6383.721 & 5.55 & -2.14 & 17.6 \\
\hline
Ni I  & 28  & 6767.780 & 1.83 & -2.17 & 11.6 \\
  &  & 5035.370 & 3.63 & 0.29 & 35.1 \\
  &  & 6772.320 & 3.66 & -0.98 & 5.2 \\
  &  & 5694.990 & 4.09 & -0.61 & 5.1 \\
\hline
Zn I  & 30  & 4810.540 & 4.08 & -0.17 & 72.5 (SS)\\
\hline
Y II  & 39  & 5289.815 & 1.03 & -1.85 & 144.7 \\
  &  & 5728.890 & 1.84 & -1.12 & 149.4 \\
\hline
Zr II  & 40  & 5124.982 & 1.53 & -1.50 & 117.2 \\
  &  & 5477.822 & 1.83 & -1.40 & 86.2 \\
\hline
La II  & 57  & 5482.268 & 0.00 & -2.06 & 158.9 \\
  &  & 5712.391 & 0.17 & -1.96 & 147.7 \\
  &  & 5936.210 & 0.17 & -2.07 & 118.5 \\
  &  & 5880.633 & 0.24 & -1.92 & 158.3 \\
  &  & 6146.523 & 0.24 & -2.47 & 83.2 \\
  &  & 5062.918 & 0.77 & -1.72 & 91.9 \\
  &  & 6126.075 & 1.25 & -1.19 & 120.6 \\
  &  & 5566.925 & 2.38 & -1.00 & 51.1 \\
  &  & 6399.030 & 2.65 & -0.52 & 70.2 \\
\hline
Ce II  & 58  & 5031.986 & 1.41 & -0.90 & 62.5 \\
  &  & 5613.694 & 1.42 & -1.00 & 67.6 \\
  &  & 5459.193 & 1.62 & -0.58 & 94.4 \\
  &  & 5959.688 & 1.63 & -0.84 & 47.6 \\
  &  & 6143.376 & 1.70 & -0.80 & 70.0 \\
  &  & 6098.326 & 1.77 & -0.61 & 61.3 \\
  &  & 5359.508 & 1.78 & -0.78 & 60.1 \\
  &  & 5685.836 & 1.90 & -0.43 & 54.4 \\
\hline
Pr II  & 59  & 5135.140 & 0.95 & -0.13 & 118.6 \\
  &  & 5605.642 & 0.96 & -0.65 & 71.3 \\
\hline
Nd II  & 60  & 5221.572 & 0.38 & -1.34 & 120.7 \\
  &  & 5934.738 & 0.75 & -1.39 & 109.7 \\
  &  & 6365.540 & 0.93 & -1.37 & 77.2 \\
  &  & 6031.270 & 1.28 & -0.83 & 103.8 \\
  &  & 6803.980 & 1.44 & -0.79 & 108.2 \\
  &  & 6737.763 & 1.60 & -0.78 & 93.4 \\
  &  & 6650.517 & 1.95 & -0.32 & 105.2 \\
\hline
Sm II  & 62  & 4791.580 & 0.10 & -1.24 & 111.2 \\
  &  & 6731.813 & 1.17 & -0.52 & 88.0 \\
\hline
Eu II  & 63  & 6437.640 & 1.32 & -0.27 & 87.2 (SS)\\
\hline
Gd II  & 64  & 5733.852 & 1.37 & -0.89 & 26.3 (SS)\\
\hline
Dy II  & 66  & 5169.688 & 0.10 & -1.95 & 68.3 (SS) \\
\hline
Er II  & 68  & 5229.319 & 2.60 & -0.39 & 16.5 (SS)\\
\hline
Yb II  & 70  & 5352.954 & 3.75 & -0.34 & 53.6 (SS)\\
\hline
Lu II  & 71 & 6463.107 & 1.46 & -1.05 & 126.0 (SS)\\
\hline
Hf II  & 72  & 4790.708 & 2.20 & -1.11 & 38.5 (SS)\\
\hline
W II  & 74  & 5104.432 & 2.36 & -0.91 & 46.3 (SS)\\
\hline
\end{longtable}

\end{document}